\documentclass[12pt,a4paper]{article}
\usepackage{pub}
\usepackage{draft}
\everymath{\displaystyle}

\usepackage[T1]{fontenc}
\usepackage{slashed}
\usepackage{caption}
\usepackage{subcaption} 
\usepackage{float}
\usepackage{empheq}
\usepackage{breqn}
\usepackage{comment}
\usepackage{xcolor}
\usepackage{color,soul}
\usepackage{bbold}
\usepackage{graphicx}
\usepackage{hyperref}
\usepackage{feynmp}

\newcommand*{\Comb}[2]{{}^{#1}C_{#2}}%

\def\del{\partial}

\def\eqref#1{(\ref{#1})}

\def\ket#1{|#1\rangle}
\def\bra#1{\langle #1 |}
\def\beq{\begin{equation}}
\def\eeq{\end{equation}}

\numberwithin{equation}{section}

\newcommand{\inv}{\frac{1}}
\newcommand{\half}{\frac{1}{2}}

\newcommand{\w}{\omega}

\newcommand{\eps}{\varepsilon}

\newcommand{\de}{\delta}

\renewcommand{\th}{\theta}
\newcommand{\al}{\alpha}

\renewcommand{\(}{\left(}
\renewcommand{\)}{\right)}

\renewcommand{\b}{\beta}
\renewcommand{\t}{\tau}
\renewcommand{\l}{\lambda}
\newcommand{\s}{\sigma}
\newcommand{\Si}{\Sigma}
\newcommand{\bpsi}{\bar{\psi}}

\newcommand{\bth}{\bar{\theta}}
\newcommand{\bPhi}{\bar{\Phi}}

\newcommand{\mN}{\mathcal{N}}
\newcommand{\bF}{\bar{F}}
\newcommand{\bphi}{\bar{\phi}}

\preprint{}

\title{A disorder-averaged framework for evaluating the  Witten index}

 \author[\dagger,*]{Indranil
  Halder,\note{ihalder@g.harvard.edu}}  \author[*]{Daniel L. Jafferis,\note{jafferis@g.harvard.edu}}\author[*]{Debmalya Sarkar\note{dsarkar@g.harvard.edu}}

\affiliation[*]{ Harvard University, Cambridge, MA 02138}
\affiliation[\dagger]{ University of California, Davis, CA 95616}

\abstract{We study supersymmetric indices in disordered systems, in particular an $\mathcal{N}=4$ supersymmetric Sachdev-Ye-Kitaev-type quantum mechanics. In cases where the disordered parameters do not affect the index, we explain how the exact answer can nevertheless be obtained using disorder averaged collective variables. Furthermore, when disorder averaging  is performed on multiple copies of the same theory, non-trivial coupling between them is generated. We show how the index ultimately remains factorized.}

\begin{document}

\maketitle

\section{Introduction}

 Microscopic understanding of black hole  physics started with work on  supersymmetry preserving  solutions of supergravity Einstein's  equations \cite{Strominger_1996}.\footnote{For a recent new prospective on it see \cite{Iliesiu:2021are, Boruch:2023gfn}.} Subsequently, their approach was refined using the Witten index of the dual quantum field theory corresponding to an AdS version of the gravitational system \cite{Maldacena_1999}. Over the past thirty years, significant progress has been made on this subject, including a prescription to obtain correlation functions of operators around supersymmetric black holes \cite{Lin:2022zxd}, the explanation of the mass gap between supersymmetric and thermal states \cite{Heydeman:2020hhw, Turiaci:2023wrh}, calculation of the thermal entropy of black holes from a microscopic point of view \cite{Halder:2023adw, Halder:2024gwe}\footnote{For related developments see \cite{Ahmadain:2022tew, Ahmadain:2022eso, Ahmadain:2024hgd, Ahmadain:2024uyo, Firat:2024kxq}.}, obtaining the Page curve for evaporating black holes \cite{penington2020entanglementwedgereconstructioninformation, Almheiri_2019}, and discovering the asymptotic probability distribution of evaporating near extremal black hole energy levels \cite{brown2024evaporationchargedblackholes, biggs2025followingstateevaporatingcharged}. 
 
 Remarkably, unitarity of evaporation of non-extremal black holes within the framework of effective field theory has been shown to be tied to the existence of wormholes \cite{Almheiri_2020I, Almheiri_2020II, penington2020replicawormholesblackhole}.\footnote{Also black holes are dual to field theories exhibiting chaotic spectra that feature level repulsion. The explanation of level repulsion within the gravitational theory comes from that of a spacetime wormhole \cite{saad2019semiclassicalrampsykgravity}. }  From the Hilbert space point of view, wormholes play a crucial role in defining the non-perturbative Hilbert space  and in identifying the null states that appear non-null in a perturbative description \cite{penington2020replicawormholesblackhole, Balasubramanian:2022gmo, akers2022blackholeinteriornonisometric, Gao_2022, iliesiu2024nonperturbativebulkhilbertspace}. %

 The discovery of numerous null states associated to non-perturbative implications of the gravitational constraint equations turns the problem of black hole entropy on its head - in a UV complete theory, rather than searching for microstates, one needs to find the specific null states that will correct down the (infinitely) overly numerous states of perturbative quantization. From this perspective, the replica wormhole calculations have a similar status to the Gibbons-Hawking euclidean black hole calculation of the entropy, in that they give a coarse-grained approximation to what should be a specific collection of null states in the microscopic theory. It remains an important open problem to understand what mechanism produces the fine-tuned cancellations required, in a single copy of a system, in a single theory. 

 Guided by history, and the obvious intractability of computing the exact spectrum of the $e^{S_{BH}}$ states of finite temperature black holes, we are motivated to ask whether the above phenomena can be observed for BPS black hole entropy. Then one might hope to give a complete accounting of all null states in the full string theory. Two potential scenarios are the two sided AdS$_2$ BPS black hole microstates of \cite{Lin:2022zxd} %
 that exhibit overcounting of the Bekenstein-Hawking entropy at the level of canonical quantization, and the BPS string index in AdS$_5 \times S^5$ that is subject to non-perturbative constraints from trace relations associated to giant gravitons \cite{lee2023tracerelationsopenstring, lee2024bulkthimblesdualtrace, batra2025giantgravitonsdpbraneholography}.
 
An immediate puzzle in the BPS context is that supersymmetry preserving wormholes do not contribute in the standard replica trick calculations, consistent with the exact factorization of BPS indexes \cite{Iliesiu:2021are}. However, the connection to replica wormholes could appear in explicitly averaged quantities. In the non-supersymmetric case, these are associated to Lorentzian configurations with coupled boundaries that admit horizonless eternally traversable wormhole solutions \cite{maldacena2018eternaltraversablewormhole}. 

  The understanding of non-supersymmetric wormhole-like physics from dual field theoretic point of view came from the work of \cite{Gao_2017, maldacena2018eternaltraversablewormhole}. In this set-up one couples two copies of the field theory  to produce  negative null energy in the bulk that leads to a traversable-wormhole like behavior at low temperatures. %
 Understanding the supersymmetric analog of these questions is an extremely interesting topic that we study in this paper. In particular, we discuss how to disorder average preserving supersymmetry in the $\mathcal{N}=4$ Sachdev-Ye-Kitaev  (SYK) model of the type studied in \cite{Anninos_2016, Murugan:2017eto, Biggs_2024}. After disorder averaging we get an effective action in terms of the two point function $G$ and self-energy $\Sigma$. We systematically study this effective action and show how to recover the exact value of the Witten index. 

In generic supersymmetric quantum field theories the Witten index is unchanged under  deformations of relevant couplings $C$ that do not change the asymptotics of the super-potential. Hence averaging over such coupling does not change the value of the index either. However such averaging might produce non-trivially disordered theories. Recovering the exact value of the Witten index from the disordered field theory is an important open problem that we solve in this paper. More, interestingly when we take product of two copies  of the index of the same theory and average over the common value of the coupling $C$ it couples two copies of the theory although the final result must be exactly factorized. In addition we can turn on further supersymmetry preserving coupling between two copies without altering the index such that  decoupled saddle point solutions are not allowed. Getting the factorized answer from this coupled representation for the product of the index is a subtle question.

This paper examines a concrete example of the above phenomena in the $\mathcal{N}=4$ SYK model. In this model the exact value of the index does not depend on the precise value of the  SYK coupling, which plays the role of the coupling constant $C$ mentioned above. After averaging over the SYK coupling with a Gaussian measure, we can recast the Witten index in terms of bi-local fields $G, \Sigma$ and their complex conjugates. The resulting theory can be studied analytically when the number of superfields are taken large. We find that the bi-local theory can be organized in terms of $G_{\phi}^{av}\equiv g$, the constant mode of the two point function of the scalar field in the super-multiplet. 

We solve the classical large $N$ equations for all fields except $g$, obtaining an effective action for $g$ including leading, sub-leading and subsub-leading orders, and show that the saddle point for $g$ is at infinity. The fact that very large values of $g$ contribute significantly despite the non-trivial super-potential of the underlying model is a feature of disorder averaging. In particular, the SYK couplings are integrated over a domain centered around vanishing coupling, a region in parameter space where the extrema of the super-potential can run off to infinity. After disorder averaging, the effective potential for $g$ becomes flat at infinity in field space. In terms of the basic fields, the saddle point value of $g$ is determined by the loop corrections. Due to the cancellation of bosons and fermions this saddle point value of $g$ at leading order in large $N$ diverges. We find, to the leading order in large $N$, effective action of the bi-local fields is flat as $g \to \infty$, and the $O(1)$ correction to the $O(N)$ classical on-shell action looks divergent in this analysis.
We argue, based on an $N=1$ explicit calculation, that it is rendered finite by $1/N$ corrections to the $G$-$\Sigma$ formalism, associated to including bi-local fields that encode correlations between different fields in the super-multiplet.

We show that the result for the index comes only from zero frequency modes of the bi-local fields, and all non-zero frequency contributions to the on-shell action cancel out due to supersymmetry.  This is in sharp contrast to the thermal entropy calculation for the model as in \cite{Biggs_2024} where the contribution comes from non-zero frequency modes. Up to the subsub-leading order that we study in this paper, we  recover the exact value of the Witten index from a calculation performed at the leading order at large $N$. We conjecture that this result will remain valid to all orders $g$. Next we analyze the product of two copies of the index. We explicitly turn on a diagonal supersymmetry preserving left-right coupling $\lambda$ and average over the common SYK coupling $C$ along with $\lambda$. In the resulting effective action $G_{RR}, G_{LL}$ couple with each other non-trivially.
For a fixed $g$, we look for `non-wormhole' like connected saddle point solutions with $G_{RL}=G_{LR}=0$. 
We identify a contributing saddle point which is left-right symmetric   and preserves diagonal supersymmetry.  We show that the path integral is dominated by the region of large $g$ where the effect of $\lambda$ is negligible. Since the saddle point does not have non-zero left right coupling we conclude that our results are compatible with   the gravitational analysis of \cite{Iliesiu:2021are}.

\section{Review of generalized Sachdev-Ye-Kitaev model}

In this paper we analyze the supersymmetric Sachdev-Ye-Kitaev type model studied in \cite{Anninos_2016, Murugan:2017eto, Biggs_2024}. These models lack a Schwarzian sector at low energies and  thus differ from conventional SYK \cite{Polchinski_2016, Maldacena_2016,maldacena2016conformalsymmetrybreakingdimensional, Fu_2017} or tensor models \cite{Witten:2016iux, Klebanov_2017, Choudhury:2017tax}. However, the supersymmetric index of these models is simple to study. In this section we introduce these models and in the next section focus on the study of the Witten index.

\subsection{$\mN =2$ supersymmetric SYK model}

$\mN =2$ superspace in one dimension includes anticommuting supercoordiantes $\th^\al, \al = 1, 2$ along with the bosonic time coordinate $\tau$. We switch between covariant and contravariant indices using the antisymmetric symbol $\eps_{\al\b}$, $\eps_{12} = \eps^{21}=1$. By convention, we contract indices from upper left to lower right and define $\th^2 =\th^\al\th_\al$.  In terms of these coordinates, the  scalar superfield is given by
\be 
\Phi(\t, \th) = \phi(\t) + i\th^\al\psi_\al(\t) -\frac{1}{2}\th^2 F(\t), 
\ee 
The components $\phi, \psi, F$ are respectively a real scalar, a Majorana fermion and an auxiliary field. For the purpose of defining a supersymmetry invariant action, it will be useful to define the covariant derivative on the superspace 
\be 
D_\al = \frac{\del}{\del\th^\al} + \th^\al\frac{\del}{\del \t},
\ee
The Euclidean action for  $\mN=2$ supersymmetric SYK model is given by\footnote{Here  we have chosen the measure such that $\int d^2\th\, \th^2 = 1 $.} 
\be 
\begin{split}\label{N2action}
    S &= \int d\t d^2\th \Big[ -\frac{i}{4}\eps^{\al\b}D_\al\Phi^iD_\b\Phi^i + C_{ijk}\Phi^i\Phi^j\Phi^k \Big]\\
    &= \int d\t \Big[\half \psi^i_\al \dot{\psi}^i_\al + \half(F^i)^2 + \half(\dot{\phi}^i)^2 + 3iC_{ijk}\big( F^i\phi^j\phi^k + \eps^{\al\b}\psi^i_\al\psi^j_\b\phi^k \big) \Big].
\end{split}
\ee 
The  $N$ superfields $\Phi^i, i\in 1, 2, \dots, N$, and the SYK coupling constant  $C_{ijk}$ is  real and symmetric in all the indices. We can also generalize the potential to 
\be
C_{ijk}\Phi^i\Phi^j\Phi^k \to C_{i_1 \dots i_p}\Phi^{i_1} \dots \Phi^{i_p},
\ee 
where $C_{i_1\dots i_p}$ are real and symmetric as before. Unless otherwise stated, we will work with the $p>2$ model in this paper.

\subsection{$\mathcal{N}=4$ supersymmetric SYK model}
 $\mN=4$ superspace consists of two anticommuting complex superspace coordinates $\th^\al$ as well as their complex conjugate $\bth_\al$. We  define the covariant derivatives by
\be 
D_\al = \frac{\del}{\del\th^\al} + \bth_\al\frac{\del}{\del\t} \ ; \ \bar{D}^\al = \frac{\del}{\del\bth_\al} + \th^\al\frac{\del}{\del\t},
\ee 
They satisfy 
\be 
\{D_\al, D_\b\} = \{ \bar{D}^\al, \bar{D}^\b \} = 0 \ ; \ \{ D_\al, \bar{D}^\b \} = 2\de^\b_\al \del_\t.
\ee 
This time we work with complex chiral (and antichiral) superfields $\Phi^i$ and $\bPhi^i$ that satisfy $\bar{D}^\al \Phi^i = D_\al\bPhi^i =0$. They can be expanded as 
\be 
\begin{split}
    \Phi(\th^\al, y) &= \phi(y)+ \sqrt{2}\th^\al \psi_\al(y)+ \th^2 F(y),  \qquad y = \t + \th^\al\bth_\al,\\
    \bPhi(\bth_\al, \bar{y}) &= \bphi(\bar{y})- \sqrt{2}\bth_\al \bpsi^\al(\bar{y})+ \bth^2 \bar{F}(y),  \qquad \bar{y} = \t - \th^\al\bth_\al.
\end{split}
\ee 
In terms of these fields, the Euclidean action for the $\mN =4$ SYK model is written as 
\be 
\begin{split}\label{N4action}
    S &= \int d\t d^4\th \bPhi^i\Phi^i + i\int d\t\Big( \int d^2\th C_{ijk}\Phi^i\Phi^j\Phi^k + \int d^2\bth \bar{C}_{ijk}\bPhi^i\bPhi^j\bPhi^k \Big)\\
    &= \int d\t \Bigg[ \dot{\bphi}^i \dot{\phi}^i + \bpsi^{i\al} \dot{\psi}^i_\al + \bar{F}^iF^i + 3i\Big(C_{ijk} (F^i\phi^j\phi^k + \eps^{\al\b}\psi^i_\al\psi^j_\b\phi^k)  + c.c\Big)\Bigg]
\end{split}
\ee 
Now the coupling constants $C_{ijk}$ are complex numbers  and symmetric in the indices. Similarly to the $\mN=2$ case, we  generalize to 
\be 
C_{ijk}\Phi^i\Phi^j\Phi^k \to C_{i_1 \dots i_p}\Phi^{i_1} \dots \Phi^{i_p}.
\ee 
and restrict our study to $p>2$ models.

\subsection{Relation between $\mN = 2$ and $\mN = 4$ models}
One important observation is, the $\mN=4$ model with $N$ chiral-antichiral multiplets is a special case of a $\mN=2$ theory, with $2N$  scalar multiplets. To see this, let's expand the complex fields and coupling constants in real and imaginary parts
\be
\begin{split}
    \phi_{(\mN=4)}^i(\t) \equiv \inv{\sqrt{2}}(\phi^i + i\phi^{i+N}) \ &, \ \bphi_{\mN=4}^i(\t) \equiv \inv{\sqrt{2}}(\phi^i - i\phi^{i+N}), \quad \dots \\
    C^{(\mN=4)}_{ijk} \equiv \inv{\sqrt{2}}(C^r_{ijk} + iC^i_{ijk}) \ &, \ \bar{C}^{\mN=4}_{ijk} \equiv \inv{\sqrt{2}}(C^r_{ijk} - iC^i_{ijk})
\end{split}
\ee 
So effectively we have broken up the $N$ complex fields $\{\phi^i_{(\mN=4)}, \psi^i_{\al(\mN=4)}, F^i_{(\mN=4)}: i=1, \dots, N\}$ to $2N$ real fields $\{\phi^I, \psi^I_\al, F^I_\al: I=1\dots, 2N\}$. With this relabeling, the kinetic terms of (\ref{N4action}) turn into the kinetic terms of (\ref{N2action}), with twice the number of fields. Furthermore, the interaction term of (\ref{N4action}) can be written down as the interaction of (\ref{N2action}), 
\be 
3i\sum_{I, J, K=1}^{2N}\tilde{C}_{IJK}(F^I\phi^J\phi^K + \eps^{\al\b}\psi^I_\al\psi^J_\b\phi^K)
\ee 
Where the interaction coefficients $\tilde{C}_{IJK}$ are related to the $\mN=4$ coefficients $C^{(\mN=4)}_{ijk} = \inv{\sqrt{2}}(C^r_{ijk} + iC^i_{ijk})$ by
\be 
\begin{split}
    &\tilde{C}_{ijk} = \inv{2}C^r_{ijk} \ ; \ \tilde{C}_{(i+N)(j+N)(k+N)} = \inv{2}C^i_{ijk}\\
    &\tilde{C}_{i(j+N)(k+N)} = \tilde{C}_{(i+N)j(k+N)} = \tilde{C}_{(i+N)(j+N)k} = -\inv{2}C^r_{ijk}\\
    &\tilde{C}_{ij(k+N)} = \tilde{C}_{i(j+N)k} = \tilde{C}_{(i+N)jk}= -\inv{2}C^i_{ijk},
\end{split}
\ee 
for $i, j, k = 1, \dots, N$. Similar statements remain valid for generic value of $p$.

\section{Disorder averaging without changing the Witten index}

 In this section we will present framework for recovering the large $N$ limit of the Witten index based on disorder averaged theory. Before beginning that analysis, we briefly review standard results on the exact index at finite $N$.
The Witten index of $\mN =2$ supersymmetric quantum mechanics (obtained after integrating out the auxillary field)
\be 
S = \int d\t \Big[ \half \psi^i_\al \dot{\psi}^i_\al + \half(\dot{\phi}^i\dot{\phi}^i)  + 3iC_{ijk}\eps^{\al\b}\psi^i_\al\psi^j_\b\phi^k + \frac{1}{2}\sum_{i=1}^N \big[\del_i W(\phi)^2\big] \Big]
\ee 
can be obtained exactly from the  formula given below (see  Appendix \ref{appW} for a review based on a `high temperature' calculation)
\be 
\begin{split}
    \mathcal{I}
    =&  \sum_{\del_iW(\{\phi^i_*\})=0,\phi^i_* \in \mathbb{R}} sgn\Big(det(\del_i\del_jW(\phi_*))\Big)
\end{split}
\ee 
For the  $\mathcal{N}=4$ SYK model with $N$ complex superfields the index can be calculated exactly and it is given by \cite{Biggs_2024}
\be
\mathcal{I}=(p-1)^{N}
\ee
The formula does not depend on the precise value of the SYK coupling $C$ since in this case it is possible to pair up $2N$ $\mathcal{N}=2$ real fields, i.e., $\phi_*$ s, to form $N$ complex fields each of which is a root of a polynomial of degree $(p-1)$ and there are $N$ such polynomial equations. 
On the other hand the value of the index for generic $\mathcal{N}=2$ theory depends on $C$ in a complicated way. This is consistent because the asymptotic behavior of generic  $\mathcal{N}=2$ theory does not posses the holomorphic structure needed for it to be a $\mathcal{N}=4$ theory.

\subsection{Warm-up:  $\mN =2$ supersymmetric SYK model}

In this sub-section, as a warm-up, we discuss the disorder averaged  $\mN =2$ supersymmetric SYK model. We will proceed in the usual way by averaging over SYK couplings and finding a description in terms of bi-local fields $G,\Sigma$. As mentioned previously, in this model the supersymmetric index depends on the SYK coupling non-trivially. Hence averaging over SYK coupling does not retain the index. Nevertheless we will find the leading order in large $N$ saddle point value of the averaged quantity. Methods similar to this sub-section will be applicable later in the context of the $\mN =4$ supersymmetric SYK model.

We perform a disorder average over a Gaussian distribution for the couplings with variance 
\be 
\langle C_{i_1 \dots i_p}C_{i'_1 \dots i'_p}\rangle_D = \frac{J^p}{p! p (N)^{p-1}}\delta_{i_1i'_1}\dots \delta_{i_p i'_p}
\ee 
We are interested in computing the index $\mathcal{I}$ of the supersymmetric quantum mechanics via a Euclidean path integral.  Around the Euclidean time circle of periodicity $\b$,   periodic boundary conditions are imposed on all the bosonic and fermionic fields. In other words, in the Fourier space the fields are given by
\be \label{fieldfourier}
\phi^i(\t) = \sum_{n=-\infty}^{\infty}\phi^i_n e^{i\w\t} \ ; \ \psi^i_\al(\t) = \sum_{n=-\infty}^{\infty}\psi^i_{\al,n} e^{i\w\t} \ ; \ F^i(\t) = \sum_{n=-\infty}^{\infty}F^i_{n} e^{i\w\t}, 
\ee 
with $\w=2n\pi/\b$. With this in mind, the disorder averaged index of the theory (\ref{N2action}) can be written as 
\be 
\begin{split}
    \langle \mathcal{I}\rangle_D &=  \int D\Phi \exp\Big[ \int dX \frac{i}{4} \eps^{\al\b}D_\al\Phi^i D_\b\Phi^i +\frac{J^p}{2pN^{p-1}} \int dX dX' \Phi^{i_1}\dots\Phi^{i_p}(X)\Phi^{i_1}\dots\Phi^{i_p}(X') \Big]
\end{split}
\ee 
where we use the short-hand $X \equiv (\t, \th^\al)$. To deal with the bilocal interaction term, we introduce the bilocal superfield $G(X, X') = \inv{N}\sum_i \Phi^i(X)\Phi^i(X')$, and integrate in 
\be 
\begin{split}
    1 &= \int DG \de(G(X, X') - \inv{N}\sum_i \Phi^i(X)\Phi^i(X'))\\
    &= \mathcal{K}\int DG D\Sigma \exp\Big[ -\frac{N}{2}\int dX dX'\Si(X, X')\Big(G(X, X') - \inv{N}\sum_i\Phi^i(X)\Phi^i(X')\Big) \Big],\\
\end{split}
\ee 
where $\mathcal{K}$ is a normalisation constant which will not play any important role, so we will omit it from the expressions from now. With this insertion, the partition function can be written as 
\be 
\begin{split}
   \langle \mathcal{I}\rangle_D&= \int D\Phi DG D\Sigma \exp\Bigg[ \int dX dX'\Big(-\Phi^i(X)(\frac{i}{4}\de(X-X')D^2 - \half \Sigma(X,X'))\Phi^i(X')\\
    &\hspace{50pt}+ \frac{J^pN}{2p}G(X, X')^p + N \Sigma(X, X')G(X, X') \Big)\Bigg]
\end{split}
\ee 
We  integrate out the scalar superfield, and get the following effective action
\be 
\begin{split}
   \langle \mathcal{I}\rangle_D&= \int DG D\Sigma \exp\Bigg[ N\Big( \log Sdet(\frac{i}{4}\de(X-X')D^2 - \half \Sigma(X,X'))\\
    &\hspace{50pt}+ \int dX dX' (\frac{J^p}{2p}G(X, X')^p +  \Sigma(X, X')G(X, X')) \Big) \Bigg],
\end{split}
\ee 
Here $Sdet$ means a super-determinant, whose expression (in terms of component fields) is given in (\ref{N2index}). Now we can write down the bilocal superfields in component form. In general the superfield $G(X, X')$ will have diagonal components like $G_{\phi\phi}, G_{FF}, G_{\psi\psi}$ as well as off-diagonal components like $G_{\phi F}, G_{\phi\psi}$
\be 
G_{\phi\phi}(\t, \t') = \inv{N}\sum_i\langle\phi^i(\t)\phi^i(\t') \rangle \, , \dots , \ G_{\phi F}(\t, \t') = \inv{N}\sum_i\langle\phi^i(\t) F^i(\t') \rangle
\ee 
 Since the off-diagonal expectation values are zero at leading order in  large $N$, we can just use the diagonal components of the bilocal fields to do the large $N$ analysis. So we expand as 
\be 
\begin{split}
    G(X_1, X_2) &= G_{\phi\phi}(\t_{12}) + \th_1^\al\th_2^\al G_{\psi\psi}(\t_{12}) - \th_1^2\th_2^2G_{FF}(\t_{12})\\
    \Sigma(X_1, X_2) &= -\Sigma_{FF}(\t_{12}) + \th_1^\al\th_2^\al \Sigma_{\psi\psi}(\t_{12}) + \th_1^2\th_2^2\Sigma_{\phi\phi}(\t_{12}),
\end{split}
\ee 
where we have used the translation symmetry in $\t$ to express the bilocal fields as functions of $\t_{12}=\tau_1-\tau_2$. In terms of the components fields, we can write the effective action for the index to the leading order in large $N$ as 
\be 
\begin{split}\label{N2index}
\frac{S_{eff}}{N}
    &= -\half \log det(-\de(\t - \t')\del_\t^2 - \Si_\phi) - \half \log det(\de(\t - \t') - \Si_F) + \log det(\de(\t - \t')\del_\t - \Si_\psi)\\
    &\quad -\int d\t d\t' \Big( \Si_\psi(\t-\t') G_\psi(\t-\t') + \half \Si_\phi(\t-\t') G_\phi(\t-\t') + \half \Si_F(\t-\t') G_F(\t-\t')\\
    &\qquad+ \frac{J^p}{2}G_F(\t-\t') G_\phi^{p-1}(\t-\t') - \frac{(p-1)J^p}{2}G_\psi^2(\t-\t')G_\phi^{p-2}(\t-\t') \Big)
\end{split}
\ee 
and reliably study the classical dynamics of \eqref{N2index} through saddle point analysis. 
Similarly to (\ref{fieldfourier}), we can expand the bilocal fields in terms of Fourier modes, 
\be 
G_A(\t) = \sum_{n=-\infty}^{\infty} G_A(\w)e^{i\w\t}, \qquad \w = \frac{2\pi n}{\b}
\ee 
and similarly for $\Sigma_A(\t)$. In terms of these frequency modes, the index is
\be 
\begin{split}
      \frac{S_{eff}}{N} &=  \sum_{\w} \Big[ \log(-i\w-\Si_\psi(\w)) - \half \log(\w^2 - \Si_\phi(\w)) - \half \log(1-\Si_F(\w)) \Big] \\
    &+ \int_0^\b d\t d\t' \Big[ -\half G_{\phi}(\t-\t')\Si_{\phi}(\t-\t') - \half G_F(\t-\t') \Si_F(\t-\t') - G_\psi(\t-\t') \Si_\psi(\t-\t')\\& - \half G_F(\t-\t') G_\phi^{p-1}(\t-\t') + \frac{(p-1)}{2}G_\psi(\t-\t')^2 G_\phi(\t-\t')^{p-2} \Big]
\end{split}
\ee 
 We can simplify this expression of the effective action  using spin-statistics of the fields $G_\psi(-\w) = -  G_\psi(\w),\quad G_\phi(-\w) =  G_\phi(\w),\quad G_F(-\w) =  G_F(\w)$ as follows
\be 
\begin{split}
     \frac{S_{eff}}{N} &= \sum_{\w} \Big[ \log(-i\w-\Si_\psi(\w)) - \half \log(\w^2 - \Si_\phi(\w)) - \half \log(1-\Si_F(\w))\\
    &\hspace{30pt} -\half G_{\phi}(\w)\Si_{\phi}(\w) - \half G_F(\w)\Si_F(\w) +  G_\psi(\w)\Si_\psi(\w)\Big] \\
    &\hspace{0pt} + \int_0^\b d\t d\t' \Big[  - \half G_F(\t-\t') G_\phi^{p-1}(\t-\t') + \frac{(p-1)}{2}G_\psi(\t-\t')^2 G_\phi(\t-\t')^{p-2} \Big]
\end{split}
\ee 
The equation of motion  for the $\Si_A$ fields gives the usual relation between two-point funtion and self-energy:
\be 
0 = \inv{-i\w - \Si_\psi(\w)} -  G_\psi(\w) = \inv{\w^2 - \Si_\phi(\w)} - G_\phi(\w) = \inv{1-\Si_F(\w)}-G_F(\w)
\ee 
Plugging these in, we get the effective action in terms of the two point function alone
\be 
\begin{split}
      \frac{S_{eff}}{N} &= \sum_{\w}\Big[ -\log(  G_\psi(\w)) + \half \log(G_{\phi}(\w)G_F(\w)) + (\half - \frac{\w^2}{2}G_\phi(\w)) + (\half - \frac{G_F(\w)}{2})\\ &+ (-i\w G_\psi(\w) - 1)  \Big] 
    + \int_{0}^\b d\t d\t' \Big[ -\half  G_F(\t) G_\phi^{p-1}(\t') + \frac{(p-1)}{2}G_\psi^2(\t) G_\phi^{p-2}(\t') \Big]\\
    &= \sum_{\w}\Big[ -\log(  G_\psi(\w)) + \half \log(G_{\phi}(\w)G_F(\w)) - \frac{\w^2}{2}G_\phi(\w)  - \frac{G_F(\w)}{2} + i\w G_\psi(\w)   \Big] \\
    & + \int_{0}^\b d\t d\t' \Big[ -\half  G_F(\t-\t') G_\phi^{p-1}(\t-\t') + \frac{(p-1)}{2}G_\psi^2(\t-\t') G_\phi^{p-2}(\t-\t') \Big]
\end{split}
\ee 
Now we expand the fields in position space as 
\be 
G_\phi (\t)= G^{av}_\phi + \de G_\phi(\t) \ ; \ G_\phi(\t) = G^{av}_F + \de G_F(\t) \ ; \ G_\phi (\t)= G^{av}_\psi + \de G_\psi(\t)
\ee 
Where $A^{av}$ quantities are the constant, or zero frequency part of the field. They are related by 
\be 
G_A(\w =0) = \int d\t G_A(\t) e^{i\w \t}|_{\w=0} = \b G^{av}_A
\ee 
For now we will keep all of $G_A(\omega=0)$ arbitrary and path integrate all the modes except  $G_\phi(\omega=0)$. This way we will generate an effective action for the path integral over  $G_\phi(\omega=0)$. Using the effective action, we will show that the saddle point value of $G_\phi(\omega=0)$  is large and it indeed preserves supersymmetry. For all the modes except $G_\phi(\omega=0)$, $G_\phi(\omega=0)$ acts as a coupling constant, so it will be convenient to define $g\equiv G_{\phi}^{av}$.  We will see that the effective energy scale of our theory is controlled by ${g^{p/2-1}}=1/l$. This suggests that it would be useful to organize our perturbation theory around $l \to 0$. 
To proceed further we follow the footsteps of \cite{Biggs_2024} and assume $\delta G_A$ are small and perform the path integration order by order in $\delta G_A$\footnote{ We find it convenient to use the following notation in the expansion: $(p-i+1)_{i-1} =  (p-1)(p-2)\dots (p-i+1)$, and we have used the shorthand $g \equiv G^{av}_{\phi}$.} 
\be 
\begin{split}
    &\int d\t d\t' \Big[ -\half  G_F(\t-\t') G_\phi^{p-1}(\t-\t') + \frac{(p-1)}{2}G_\psi^2(\t-\t') G_\phi^{p-2}(\t-\t') \Big] \\
    =& \int d\t d\t' \half\Big[  - G^{av}_F g^{p-1}+(p-1)g^{p-2} (G^{av}_\psi)^2 + (p-1)g^{p-2} \de G_\psi^2 - (p-1)g^{p-2} \de G_F \de G_\phi \\
    &\hspace*{2cm}  - \frac{(p-2)_2g^{p-3}}{2}G^{av}_F \de G_\phi^2 + 2(p-2)_2g^{p-3} G^{av}_\psi \de G_\psi \de G_\phi + (p-2)_2g^{p-3}\de G_\psi^2\de G_\phi\\
    & \hspace*{2cm} - \frac{(p-2)_2g^{p-3}}{2} \de G_F \de G_\phi^2 - \frac{(p-3)_3g^{p-4}}{6}G^{av}_F \de G_\phi^3+ (p-3)_3g^{p-4} G^{av}_\psi \de G_\psi \de G_\phi^2\\
    &\hspace*{2cm}+ \frac{(p-3)_3g^{p-4}}{6}(3(G^{av}_\psi)^2\de G_\phi^2+ 3\de G_\psi^2 \de G_\phi^2- \de G_F \de G_\phi^3) +\dots \Big]
\end{split}
\ee 
Note that each factor is subleading to the previous one by a factor of $g \equiv G^{av}_\phi$. To the quadratic order in  $\delta G_A$ and leading order in $g$  we get (we will work under the assumption that higher powers of $g$ are sub-leading. A complete justification for this assumption will appear later.)
\be 
\begin{split}
     \frac{S_{eff}}{N}&= -\frac{\b}{2} G_F(\w=0) g^{p-1} + \frac{(p-1)g^{p-2}}{2} G_\psi(\w=0)^2\\
     &\hspace{30pt}+ \frac{(p-1)g^{p-2}}{2}\sum_{\w \neq 0} \Big[ -\de G_F(\w)\de G_\phi(\w) -\de G_\psi(\w)^2 \Big]\\
    &\hspace{30pt}+ \sum_{\w}\Big[ -\log(G_\psi(\w)) + \half \log(G_{\phi}(\w)G_F(\w)) - \frac{\w^2}{2}G_\phi(\w)  - \frac{G_F(\w)}{2} - i\w G_\psi(\w) \Big]\\
\end{split}
\ee 
The equation of motion for $G_F(\w=0), G_\psi(\w=0)$ determines these in terms of $G_\phi^{av}$:
\be 
\inv{G_F(\w=0)} -1 -\b (G^{av}_\phi)^{p-1} =0 \implies G_F(\w=0) = \inv{1+ \b (G^{av}_\phi)^{p-1}}
\ee 
\be 
(p-1)g^{p-2} G_\psi (\w=0) - \inv{G_\psi(\w=0)}=0 \implies G_\psi(\w=0) = \inv{\sqrt{(p-1)(G^{av}_\phi)^{p-2}}}
\ee 
The equation of motion for $G_\phi(\w=0)$ is given as follows:
\be
\frac{1}{2G_\phi(\w=0)}-\frac{1}{2} (p-1) \frac{\b^{2-p}(G_\phi(\w=0))^{p-2}}{1+ \b^{2-p} (G_\phi(\w=0))^{p-1}}+\frac{(p-2)}{2G_\phi(\w=0)} + \text{($\w\neq0$ contributions)}=0
\ee
Here we have plugged back the solution for $G_F(\w=0), G_\psi(\w=0)$. The first term above comes from the one loop effect of $\phi$. The second, third term are due to the interaction with $F, \psi$ as dictated by supersymmetry. If we ignore the effect of non-zero $\w$ modes and focus only on bosonic interactions then the saddle point value of  $G_\phi(\w=0)$ could be either $\infty$ or some finite value dependent on $p$. On the other hand, when  the interaction with the fermion is included the only allowed solution is $G_\phi(\w=0)=\infty$. This is another manifestation of the fact that it is supersymmetry which makes the saddle point value of $G_\phi(\w=0)$ large. An even more direct way of seeing this is as follows: in the integral representation of 
$ \langle \mathcal{I}\rangle_D$ ignore contributions of non-zero modes. Then the bosonic saddle point of $G_F(\w=0)$ is such that the exponential term in $ G_\phi(\w=0)$  cancels out up to a constant piece and the saddle point is solely determined by the power law terms coming from the one loop effects in $F, \phi$. When $\psi$ is included again the exponential term in $ G_\phi(\w=0)$ cancels out up to a constant, but now the one loop effect of $\psi$ shifts the saddle point value of $G_\phi(\w=0)$ to infinity. Note that $G_\phi(\w=0)$ depends non-trivially on the details of the coupling constants $C_{ij\dots}$ and we see that the disorder averaged value of $G_\phi(\w=0)$ diverges. This does not necessarily imply that $G_\phi(\w=0)$ before disorder averaging was not bounded. In fact we expect  $G_\phi(\w=0)$ to be bounded before disorder averaging for generic values of   $C_{ij\dots}$. For more discussion of related questions see appendix \ref{appA}. 

 Nonzero mode equations of motions for $\w>0$ are 
\be 
\begin{split}
    -\inv{\delta G_\psi(\w)} - i\w - (p-1)&g^{p-2} \delta G_\psi(\w) = 0 \ ; \ \inv{\delta G_\phi(\w)} - \w^2 - (p-1)g^{p-2} \delta G_F(\w) = 0 \\
    &\inv{\delta G_F(\w)} - 1 - (p-1)g^{p-2} \delta G_\phi(\w) = 0
\end{split}
\ee 
For non-zero modes the leading order solutions will be denoted as  $\delta G_A(\w)=Q_A(\w)$.  These are given by
\be 
\begin{split}
Q_\psi &= sgn(\w) \inv{\sqrt{(p-1)g^{p-2}}} i H(\w), \ \ \ H(\w) = \Big[ -\frac{|\w|}{2\sqrt{(p-1)g^{p-2}}} + \sqrt{1 + \frac{\w^2}{4(p-1)g^{p-2}}} \Big]\\
Q_F &= \inv{\sqrt{(p-1)g^{p-2}}}|\w| H(\w)\\
Q_\phi &= \inv{\sqrt{(p-1)g^{p-2}}}\inv{|\w|}H(\w)
\end{split}
\ee 
Substituting the solutions, the index takes the form 
\be 
\begin{split}
     \frac{S_{eff}}{N} &= -\frac{\b}{2} G_F(\w=0)g^{p-1} + \frac{(p-1)g^{p-2}}{2}G_\psi(\w=0)^2\\
    &\hspace{30pt}+ \sum_{\w}\Big[ -\log(G_\psi(\w)) + \half \log(G_{\phi}(\w)G_F(\w)) - \frac{\w^2}{2}G_\phi(\w)  - \frac{G_F(\w)}{2} - i\w G_\psi(\w) \Big]\\
    &\hspace{30pt} + \frac{(p-1)g^{p-2}}{2}\sum_{\w \neq 0} \Big[ -\de G_F(\w)\de G_\phi(\w) - \de G_\psi(\w)^2 \Big]\\
    &= -\half\frac{\b g^{p-1}}{1+ \b g^{p-1}} + \half  - \log(\inv{\sqrt{(p-1)g^{p-2}}}) + \half \log(\frac{\b g}{1 + \b (g)^{p-1}}) - \half\inv{1+\b g^{p-1}} \\
    &\hspace{30pt} +2\sum_{\w > 0} \Big[ \half \log (\frac{H(\w) H(\w)}{H(w)^2}) + \inv{\sqrt{(p-1)g^{p-2}}} (\w - \w/2 - \w/2)H(\w)\\
    &\hspace{40pt}+ \frac{(p-1)g^{p-2}}{2}(-\inv{(p-1)g^{p-2}} H(\w)^2 + \inv{(p-1)g^{p-2}}H(\w)^2) \Big]\\
    &= \half\Big[ 1 - \frac{\b g^{p-1}}{1+ \b g^{p-1}} - \inv{1+\b g^{p-1}} + \log(\frac{(p-1)\b g^{p-1}}{1 + \b g^{p-1}}) \Big] + 0\\
    &= \half \log(p-1) + \half \log(\frac{\b (G^{av}_\phi)^{p-1}}{1+\b (G^{av}_\phi)^{p-1}})
\end{split}
\ee 
We see remarkable cancellation due to supersymmetry in above solutions
\be
& G_F(\t)=-\partial_{\t}^2 G_\phi(\t), \quad G_\psi(\t)=-\partial_{\t} G_\phi(\t) \\
\implies & G_F(\w)=\w^2 G_\phi(\w), \quad G_\psi(\w)=i\w G_\phi(\w)
\ee
This is different from the thermal calculation presented in \cite{Biggs_2024}. In fact for us the entire contribution to $ \log(p-1)/2$ came from the zero frequency modes.  Now we turn to $G_\phi(\w=0)$ path integration and note from above expression that it is peaked at $G_\phi(\w=0)\to \infty$. This is perfectly consistent with the fact that supersymmetry implies the contribution from $G_\phi(\w=0)$ path integration to the index comes only from the region where $G_\phi(\w=0)\to \infty$. Hence the leading order value of the averaged index is given by \cite{auffinger2011randommatricescomplexityspin}
\be 
\begin{split}
    \frac{\log \langle \mathcal{I}\rangle_D}{N}
    &= \half \log(p-1) 
\end{split}
\ee 

\subsection{$\mN =4$ supersymmetric SYK model}

We now turn to the $\mN =4$ supersymmetric SYK model. In this model the value of the Witten index does not depend on the precise values of the SYK coupling constants. Hence we can average over them without changing the value of the index. However averaging produces an effective action in terms of bi-local fields $G, \Sigma$ and it is not entirely clear how to obtain the exact value of the index from their large $N$ classical solution. In this sub-section we solve this problem. First we show the exact value of the index is recoverable from a classical solution that is valid to the leading order  in an effective energy scale $g$ and then a present a detailed analysis up to sub-sub leading order in $g$ to show that the value of the leading order calculation is essentially unchanged. 

\subsubsection{Quadratic solution at leading order in $g$}
The effective action in terms of bi-local fields can be derived in the same way as in the $\mathcal{N}=2$ theory. We just present the result here, the interested reader can find a derivation of it in \cite{Biggs_2024} 
 \be 
\begin{split}
\frac{S_{eff}}{N} &=\sum_{\w } \Big[- \log(\w^2 - \bar{\Si}_\phi(\w)) - \log(1- \bar{\Si}_F(\w)) + 2\log(-i\w - \bar{\Si}_\psi(\w))\Big]\\
    &\hspace{30pt} - \int d\t d\t' \Big[ 2\Si_\psi \bar{G}_\psi + \Si_\phi \bar{G}_\phi + \Si_F\bar{G}_F - (p-1)G_\psi^2 G_\phi^{p-2} + G_F G_\phi^{p-1} \Big]\\
    &= \sum_{\w } \Big[- \log(\w^2 - \bar{\Si}_\phi(\w)) - \log(1- \bar{\Si}_F(\w)) + 2\log(-i\w - \bar{\Si}_\psi(\w))\Big]\\
    &\hspace{30pt} - \sum_{\w } \Big[ 2\Si_\psi(\w)G_\psi(-\w) + \Si_\phi(\w)G_\phi(-\w) + \Si_F(\w)G_F(-\w) \Big]\\
    &\hspace{30pt} + \int d\t d\t' \Big[ (p-1)G_\psi^2(\t-\t') G_\phi^{p-2}(\t-\t') - G_F(\t-\t') G_\phi^{p-1}(\t-\t') \Big]
\end{split}
\ee 
Let's explain the terms appearing: we have defined 
\be 
\begin{split}
    G_\phi(\t, \t') = \inv{N}\langle \bphi^i(\t)\phi^i(\t') \rangle \ &; \ \bar{G}_\phi(\t, \t') = G_\phi(\t', \t) = \inv{N}\langle \phi^i(\t)\bphi^i(\t') \rangle \\
    G_\psi(\t, \t') = \inv{N}\langle \bpsi^{\al i}(\t)\psi^i_{\al}(\t') \rangle \ &; \ \bar{G}_\psi(\t, \t') = -G_\psi(\t', \t) = \inv{N}\langle \psi^{\al i}(\t)\bpsi^i_{\al}(\t') \rangle\\
    G_F(\t, \t') = \inv{N}\langle \bar{F}^i(\t) F^i(\t') \rangle \ &; \ \bar{G}_F(\t, \t') = G_F(\t', \t) = \inv{N}\langle \bar{F}^i(\t) F^i(\t') \rangle
\end{split}
\ee 
Their conjugate variables are $\Si_A(\t, \t'), \bar{\Si}_A(\t, \t')$. If we make the ansatz $G_A(\t_1, \t_2) \equiv G_A(\t_{12})$, then we find the following relation among the Fourier series coefficients 
\be 
\bar{G}_\phi(\w) = G_\phi(-\w) \ ; \ \bar{G}_\psi(\w) = -G_\psi(-\w) \ ; \  \bar{G}_F(\w) = G_F(-\w) \ ; \
\ee 
Using these relations, we can replace the barred quantities  
\be 
\begin{split}
    \frac{S_{eff}}{N} &= \sum_{\w } \Big[- \log(\w^2 - {\Si}_\phi(\w)) - \log(1- {\Si}_F(\w)) + 2\log(i\w + {\Si}_\psi(\w))\Big]\\
    &\hspace{30pt} + \sum_{\w } \Big[ 2\Si_\psi(\w){G}_\psi(\w) - \Si_\phi(\w){G}_\phi(\w) - \Si_F(\w){G}_F(\w) \Big]\\
    &\hspace{10pt} + \int d\t d\t' \Big[ (p-1)G_\psi(\t-\t')^2 G_\phi(\t-\t')^{p-2} - G_F(\t-\t') G_\phi(\t-\t')^{p-1} \Big]
\end{split}
\ee 
We use the $\Si_A$ equations of motion, 
\be 
\begin{split}
    \inv{\w^2 - {\Si}_\phi(\w)} = G_\phi(\w) \ ; \ \inv{1- \Si_F(\w)} = G_F(\w) \ ; \ \inv{i\w + \Si_\psi(\w)} =- G_\psi(\w)
\end{split}
\ee 
We further assume that $G_\psi(-\w)=-G_\psi(-\w),G_\phi(-\w)=G_\phi(-\w),G_F(-\w)=G_F(-\w) $. This simplifies the expression of the action to 
\be 
\begin{split}
    \frac{S_{eff}}{N} &= \sum_{\w} \log(G_\phi(\w)G_F(\w)) - 2\log(G_\psi(\w)) +\sum_{\w}\Big[ -2i\w G_\psi(\w) - \w^2G_\phi(\w) - G_F(\w)  \Big] \\
    &+ \int d\t d\t' \Big[ (p-1)G_\psi(\t-\t')^2 G_\phi(\t-\t')^{p-2} - G_F(\t-\t') G_\phi(\t-\t')^{p-1} \Big]
\end{split}
\ee 
To proceed further we follow the footsteps of \cite{Biggs_2024} and assume $\delta G_A$ are small and expand order by order in $\delta G_A$.  
\be 
\begin{split}
    &\int d\t d\t' \Big[ -  G_F(\t-\t') G_\phi^{p-1}(\t-\t') + (p-1)G_\psi^2(\t-\t') G_\phi^{p-2}(\t-\t') \Big] \\
    =& \int d\t d\t' \Big[ (p-1)g^{p-2} (G^{av}_\psi)^2 - G^{av}_F (G^{av}_\phi)^{p-1}+ (p-1)g^{p-2} (\de G_\psi^2 -  \de G_F \de G_\phi) \\
    & 
    \quad - \frac{(p-2)_2g^{p-3}}{2}(G^{av}_F \de G_\phi^2  + \de G_F \de G_\phi^2-2 \de G_\psi^2\de G_\phi-4G^{av}_\psi \de G_\psi \de G_\phi) \\
    & \quad +\frac{(p-3)_3g^{p-4}}{6}(3(G^{av}_\psi)^2\de G_\phi^2 + 6 G^{av}_\psi \de G_\psi \de G_\phi^2 + 3\de G_\psi^2 \de G_\phi^2-G^{av}_F \de G_\phi^3 - \de G_F \de G_\phi^3)\\
    & \quad \frac{(p-4)_4g^{p-5}}{3}G^{av}_\psi \de G_\phi^3 \de G_\psi + \frac{(p-5)_5g^{p-6}}{24}(G^{av}_\psi)^2 \de G_\phi^4 - \frac{(p-4)_4g^{p-5}}{24}G^{av}_F \de G_\phi^4 + \mathcal{O}(\de G_A^5) \Big]
\end{split}
\ee 
Note that each factor is sub-leading to the previous one by a factor of $G^{av}_\phi$. 
The terms can be grouped into 
\be \label{pieces}
\begin{split}
    &\text{zero mode: } (p-1)g^{p-2} (G^{av}_\psi)^2 - G^{av}_F g^{p-1}=\mathcal{O}(1)\\
    &\text{leading quadratic: } (p-1)g^{p-2} (\de G_\psi^2 - \de G_F \de G_\phi)=\mathcal{O}(1) \\
    &\text{sub-leading quadratic: } 2(p-2)_2g^{p-3} G^{av}_\psi \de G_\psi \de G_\phi=\mathcal{O}(g^{-\frac{p}{2}})\\
    &\text{subsub-leading quadratic: }  \Big(- \frac{(p-2)_2g^{p-3}}{2} G^{av}_F +  \frac{(p-3)_3g^{p-4}}{2}(G^{av}_\psi)^2\Big) \de G_\phi^2=\mathcal{O}(g^{-p}) \\
    &\text{leading cubic: } \frac{(p-2)_2g^{p-3}}{2}(2\de G_\psi^2 \de G_\phi - \de G_F \de G_\phi^2) =\mathcal{O}(g^{-\frac{p}{2}})\\
     &\text{sub-leading cubic: }   (p-3)_3g^{p-4}G^{av}_\psi \de G_\psi \de G_\phi^2=\mathcal{O}(g^{-p}) \\
     &\text{subsub-leading cubic: }   -\frac{(p-3)_3g^{p-4}}{2}  G^{av}_F \de G_\phi^3=\mathcal{O}(g^{-\frac{3p}{2}})\\
    &\text{leading quartic: } \frac{(p-3)_3g^{p-4}}{6}(3\de G_\psi^2 \de G_\phi^2 - \de G_F \de G_\phi^3)=\mathcal{O}(g^{-p}) \\
    &\text{sub-leading quartic: } \frac{(p-4)_4g^{p-5}}{3}G^{av}_\psi \de G_\phi^3 \de G_\psi = \mathcal{O}(g^{-\frac{3p}{2}})\\
     &\text{subsub-leading quartic: } \Big(\frac{(p-5)_5g^{p-6}}{24}(G^{av}_\psi)^2  - \frac{(p-4)_4g^{p-5}}{24}G^{av}_F \Big) \de G_\phi^4 = \mathcal{O}(g^{-2p})
\end{split}
\ee 

For power counting of $g \equiv G^{av}_\phi$, we note that apart from $G^{av}_F= \mathcal{O}(1/g^{p-1})$ all of $Q_A$ and $G^{av}_\psi$ are $\mathcal{O}(g^{1-p/2})$. To the quadratic order in  $\delta G_A$ and leading order in $g$ we get (terms containing higher powers of $g$ contribute at the sub-leading order, we discuss them later in the paper)
\be 
\begin{split}
    \frac{S_{eff}}{N} &=\sum_{\w} \bigg[ \log(G_\phi(\w)G_F(\w)) - 2\log(G_\psi(\w))\bigg]+ \sum_{\w}\Big[ -2i\w G_\psi(\w) - \w^2G_\phi(\w) - G_F(\w)  \Big] \\
    & + (p-1)g^{p-2} \Bigg(G_\psi^2(\w=0) -\frac{\b g}{p-1} G_F(\w=0)  + \sum_{\w \neq 0} \Big[\delta G_\psi(-\w)\delta G_\psi(\w)  -\delta G_F(-\w)\delta G_\phi(\w)  \Big]\Bigg)
\end{split}
\ee 
The zero mode equations of motion of all the modes except $G_\phi(\omega=0)$ are given by 
\be 
G_F(\w=0) = \inv{1+\b g^{p-1}}, \quad G_\psi(\w=0)= \inv{\sqrt{(p-1)g^{p-2}}}
\ee
And then nonzero mode equations of motions for $\w>0$ are 
\be 
\begin{split}
    \inv{\delta G_\phi(\w)} - \w^2 - &(p-1)g^{p-2} \delta G_F(-\w) = \inv{\delta G_F(\w)} - 1 - (p-1)g^{p-2} \delta G_\phi(-\w) =0\\
    &-\inv{\delta G_\psi(\w)} - i\w + (p-1)g^{p-2} \delta G_\psi(-\w) = 0
\end{split}
\ee 
As in the $\mN=2$ case, these are solved in terms of the same function
\be 
\begin{split}
Q_\psi &= sgn(\w) \inv{\sqrt{(p-1)g^{p-2}}} i H(\w), \ \ \ H(\w) = \Big[ -\frac{|\w|}{2\sqrt{(p-1)g^{p-2}}} + \sqrt{1 + \frac{\w^2}{4(p-1)g^{p-2}}} \Big]\\
Q_F &= \inv{\sqrt{(p-1)g^{p-2}}}|\w| H(\w)\\
Q_\phi &= \inv{\sqrt{(p-1)g^{p-2}}}\inv{|\w|}H(\w)
\end{split}
\ee 
We evaluate the effective action on the solution,
\be 
\begin{split}
    \frac{S_{eff}}{N} &= -\frac{\b g^{p-1}}{1 + \b g^{p-1}} + 1 + \log(\frac{\b g}{1+\b g^{p-1}}) - 2 \log(\inv{\sqrt{(p-1)g^{p-2}}}) - \inv{1+\b g^{p-1}}\\
    &\hspace{20pt} \sum_{\w \neq 0} \log(\frac{H(\w )H(\w )}{H(\w )^2}) + \inv{\sqrt{(p-1)g^{p-2}}}\Big( 2|\w | H(\w ) - |\w | H(\w ) - |\w |H(\w ) \Big)\\
    &\hspace{50pt}+ \inv{(p-1)g^{p-2}}\Big( H(\w )^2 - H(\w )^2 \Big)\\
    &= \log(\frac{(p-1)\b g^{p-1}}{1+\b g^{p-1}}) \\
    &= \log(p-1) - \log(1 + \inv{\b (G^{av}_\phi)^{p-1}})
\end{split}
\ee 
Just like in $\mathcal{N}=2$ model, the path integral over $G_\phi(\w=0)$ localizes onto $G_\phi(\w=0) \to \infty$
region giving 
\be 
\begin{split}
    \frac{\log\mathcal{I}}{N} 
    &= \log(p-1) 
\end{split}
\ee 

\subsubsection{Quadratic solution  at sub-leading orders in $g$}

In this sub-section we discuss classical contributions at sub-leading orders in $g$. 
These come from both sub-leading quadratic terms and from higher order in $\delta G_A$ terms in (\ref{pieces}). As a first step we evaluate them on the quadratic solution presented before. We will see that this is sub-leading in $g$ compared to the result presented before. This  justifies ignoring corrections due to these terms in the classical solution for evaluating the first sub-leading correction to the potential for $g$. 

Recall that supersymmetry  implies
\be 
Q^\psi = -\del_\t Q^\phi \ , \ Q^F = -\del^2_\t Q^\phi
\ee 
Then the sub-leading quadratic term can be written as (up to an overall prefactor of $2(p-2)_2\,g^{p-3} \b$, see \eqref{pieces}) 
\be 
- G^{av}_\psi\int d\t Q^\phi Q^\psi =  G^{av}_\psi \int d\t Q^\phi \del_\t Q^\phi =  G^{av}_\psi \int d\t \half \del_\t((Q^\phi)^2) = 0 
\ee 
The first term in the subsub-leading quadratic piece, computed on the saddle gives, 
\be 
\begin{split}
  &  -\frac{(p-2)_2\,g^{p-3}}{2}\int d\t d\t' G^{av}_F \de G_{\phi}^{2}\\
  &= -\frac{(p-2)_2\,g^{p-3}}{2}\inv{\b(1 + \b g^{p-1})} \sum_{\w \neq 0} \inv{(p-1)g^{p-2} \w^2} H(\w)^2  \\
    & =  -\frac{(p-2)_2\,g^{p-3}}{2}\inv{\b(1 + \b g^{p-1})} \sum_{\w \neq 0} \frac{1}{((p-1)g^{p-2})^2} \( \frac{(p-1)g^{p-2}}{\w^2} + \inv{2}-\frac{1}{2} \sqrt{1 + \frac{4 (p-1)g^{p-2}}{\w^2}} \)\\
    &=  -\frac{(p-2)_2\,g^{p-3}}{2}\inv{\b(1 + \b g^{p-1})} \Big(\frac{\b^2}{2\pi^2 (p-1)g^{p-2}} \zeta(2)\\
    &\hspace{70pt}-\sum_{k=1}^{\infty}\frac{1}{ ((p-1)g^{p-2})^2}\Comb{1/2}{k} \zeta(2k)(\frac{\b\sqrt{(p-1)g^{p-2}}}{\pi})^{2k}\Big)
\end{split}
\ee 
We have the following upper bound on the sum over $k$ appearing in the expression above
\be
\sum_{k=1}^{\infty}\frac{1}{ a^2}\Comb{1/2}{k} \zeta(2k)(\frac{\b\sqrt{a}}{\pi})^{2k} \leq \sum_{k=1}^{\infty}\frac{1}{ a^2} |\Comb{1/2}{k}| \zeta(2k)(\frac{\b\sqrt{|a|}}{\pi})^{2k}\leq \frac{\b^2}{\pi^2a}\sum_{k=1}^{\infty}\zeta(2k)(\frac{\b\sqrt{|a|}}{\pi})^{2k-2}, 
\ee
where $a = (p-1)g^{p-2}$. The sum over $k$ in RHS is the famous Apéry-like formula for the zeta function valid when $x$ is  not  a
nonzero integer \cite{em/1175789759}
\be
\sum_{k=1}^{\infty}\zeta(2k)x^{2k-2}=\frac{1-\pi x \cot(\pi x)}{2x^2}
\ee
Hence the sum over $k$ contributes at $ \mathcal{O}((g^{p-2})^{-3/2})$ for large $g$, which is subleading. Thus we have
\be 
 -\frac{(p-2)_2\,g^{p-3}}{2}\int d\t d\t' G^{av}_F \de G_{\phi}^{2}\approx -\frac{(p-2)_2\,g^{p-3}}{(p-1)g^{p-2}}\frac{\b\zeta(2)}{4\pi^2(1+\b g^{p-1})}  \approx  - \frac{(p-2)}{24}\frac{ \b}{ G^{av}_\phi (1 + \b (G^{av}_\phi)^{p-1})}
\ee
Similar manipulation determines the second term in the subsub-leading quadratic piece to be
\be 
 \frac{(p-3)_3\, g^{p-4}}{2}\int d\t d\t' (G^{av}_\psi)^2 \de G_{\phi}^{2}  \approx   \frac{ (p-2)(p-3)}{24(p-1)}\frac{1}{ (G^{av}_\phi)^{p}}
\ee
On the other hand the leading and sub-leading cubic terms vanish because of supersymmetry:
\be 
& \frac{(p-2)_2\,g^{p-3}\b}{2} \int d\t (2\de G_\psi^2 \de G_\phi - \de G_F \de G_\phi^2)\\
=&\frac{(p-2)_2\,g^{p-3} \b }{2} \int d\t \Big[ 2 Q^\phi (\del_\t Q^\phi)^2 +\del_\t^2 Q^\phi (Q^\phi)^2  \Big]\\
=& \frac{(p-2)_2\,g^{p-3} \b }{2} \int d\t \del_\t \Big( \del_\t Q^\phi (Q^\phi)^2\Big) = 0
\ee
and 
\be
 -(p-3)_3\, g^{p-4} \b \int d\t G^{av}_\psi \de G_\psi \de G_\phi^2= \frac{(p-3)_3\, g^{p-4}}{3} \b G^{av}_\psi \int d\t   \partial_\tau   (Q^\phi)^3=0
\ee
In this paper we restrict our analysis up to $\mathcal{O}((G^{av}_\phi)^{-p})$. Therefore sub-sub leading cubic term is not our concern. Finally the leading quartic term is again a total derivative due to supersymmetry constraints:
\be 
\frac{(p-3)_3\, g^{p-4}}{6} \b \int d\t [3 \de G_\psi^2 - \de G_F \de G_\phi]\de G_\phi^2 = \frac{(p-3)_3\, g^{p-4}}{6} \b \int d\t \del_\t ((Q^\phi)^3 \del_\t  Q^\phi) = 0
\ee 
Putting all these together we get following effective action for $G^{av}_{\phi}$ at the classical level
\be 
\begin{split}\label{effA}
    \frac{\log\mathcal{I}}{N}
    &= \log(p-1) - \log(1 + \inv{\b g^{p-1}})\\
    & \hspace{2 cm }+\frac{ (p-2)(p-3)}{24(p-1)}\frac{1}{ g^{p}} - \frac{(p-2)}{24}\frac{ \b^2}{\b g (1 + \b g^{p-1})}+\mathcal{O}(g^{-\frac{3p}{2}})\\
    &= \log(p-1) - \inv{\b (G^{av}_\phi)^{p-1}}-\frac{ (p-2)}{12(p-1)}\frac{1}{ (G^{av}_\phi)^{p}}+\mathcal{O}((G^{av}_\phi)^{-\frac{3p}{2}})
\end{split}
\ee 
The action looks significantly different from the thermal solution discussed in \cite{Biggs_2024}. In particular we see that both the $G^{av}_\phi$ dependent terms above are of same sign, indicating that the action is maximized at $G^{av}_{\phi} \to \infty$. Setting this saddle point value to $G^{av}_\phi$ gives us the exact value of the supersymmetric  index.

\subsubsection{Corrections to quadratic solution}

So far we have considered the solution $Q_A$  to the equations of motion coming from only leading terms as $G^{av}_\phi \to \infty$. Now we correct the solution by taking into account the presence of other terms.

\textbf{Corrections to non-zero frequency modes:}
We introduce $\delta G_A=Q_A+g_A$. The equations of motion for $g_A$ are not automatically satisfied since $Q_A$ are not exact classical solutions.  Given the explicit factor of $g$ in leading quadratic term $ g (\de G_\psi^2 - \de G_F \de G_\phi)$ and the quadratic term coming from the logarithms, we conclude that the quadratic term for $g_A$ has an explicit factor of $g$. As a result, a linear term in $g_A$ on the new saddle point is of the same order as the square of the term it came from (i.e., obtained by replacing $g_A$ with $Q_A$) on the old saddle point. $Q_A$ solves the equations of motion for the zero mode and the leading quadratic term, hence those terms don't generate any linear term in $g_A$. Hence, up to order $(G^{av}_\phi)^{-p}$ that we are keeping track of in this note, the only source of linear terms are from sub-leading quadratic and leading cubic term. We analyze these terms below. Note the distinction between $g\equiv G^{av}_\phi$ and $g_A$ in the following analysis. 

The quadratic part of the action for $g_A$ (we discuss the linear terms separately) comes from the logarithms and the leading quadratic part involving $\delta G_A$. It  takes the following form
\be 
\begin{split}
    \frac{S_{eff}}{N}
    \supset & \sum_{\w\neq 0} \bigg[ \frac{g_\psi(\w)^2}{(Q^\psi(\w))^2} - \frac{g_\phi(\w)^2}{2(Q^\phi(\w))^2} -\frac{g_F(\w)^2}{2(Q^F(\w))^2} - (p-1)g^{p-2}(g_F(\w)g_\phi(-\w) -g_\psi(\w)g_\psi(-\w))\bigg]\\
    \approx& \sum_{\w \neq 0} \bigg[ -\frac{(p-1)}{2}g^{p-2}\Big( |\w|g_\phi (\w)+ \frac{g_F(\w)}{|\w|} \Big)^2 - 2(p-1)g^{p-2} g_\psi(\w)^2\\
    &\hspace{4cm}- \frac{\sqrt{(p-1)g^{p-2}}}{2|\w|}( |\w|^4g_\phi(\w)^2+g_F(\w)^2)- \sqrt{(p-1)g^{p-2}}|\w| g_\psi(\w)^2\bigg] 
\end{split}
\ee 
To obtain the second line we expanded $Q_A$ around large $g$ and kept leading and first sub-leading order terms. To diagonalize the quadratic term we perform a field re-definition:
\be 
g_\phi(\w) = \inv{|\w|}(\rho (\w)+ \gamma(\w)), \  \ g_F = |\w|(\rho (\w)- \gamma(\w))
\ee 
Plugging this back into the action we see that the non-zero quadratic term for $\gamma$ appears from the first sub-leading term above. To the leading order in $g$ the quadratic part of the action takes the following form
\be 
 \frac{S_{eff}}{N}
    \supset \sum_{\w\neq 0} \bigg[-2(p-1)g^{p-2} \rho(\w)^2 - \sqrt{(p-1)g^{p-2}}|\w|\gamma(\w)^2 - 2(p-1)g^{p-2} g_\psi(\w)^2\bigg]
\ee 
Now we turn to evaluate the linear terms in $g_A$. However before proceeding, note that spin-statistics requires
\be
g_\psi(-\w)=-g_\psi(\w), \quad g_\phi(-\w)=g_\phi(\w),\quad g_F(-\w)=g_F(\w)
\ee
An immediate consequence of this is that the linear term in $g_A$ from sub-leading quadratic term vanishes 
\be 
&-2(p-2)_2\,g^{p-3} \b \int d \tau G^{av}_\psi \de G_\psi(\t) \de G_\phi(\t) \\
&= 2(p-2)_2\,g^{p-3} \b \int d \t G^{av}_\psi (Q_\psi(\t) Q_\phi(\t) + Q_\psi(\t) g_\phi(\t) + Q_\phi(\t) g_\psi(\t) + g_\phi(\t) g_\psi(\t))\\
&= 2(p-2)_2\,g^{p-3} \b \int d \t G^{av}_\psi Q_\psi(\t) Q_\phi(\t)
\ee 
The contribution of the leading cubic term is more involved. We first focus on the cubic term involving $g_\psi$:
\be 
(p-2)_2\,g^{p-3}\b \int d\t \ 2Q^\psi(\t) g_\psi(\t) Q^\phi(\t) 
\ee
Together with the quadratic term above, the classical solution for $g_\psi$ is determined from  
\be 
4(p-1)g^{p-2} g_\psi(\t) + 2(p-2)_2\,g^{p-3}Q^\psi(\t)Q^\phi(\t)=0
\ee 
Plugging this back to the action gives
\be \label{Iferm}
\begin{split}
    \frac{S_{eff}}{N} &\supset \frac{((p-2)_2\,g^{p-3})^2}{2(p-1)g^{p-2}}\b\int d\t (Q^\psi(\t) Q^\phi(\t))^2 -\frac{((p-2)_2\,g^{p-3})^2}{(p-1)g^{p-2}}\b\int d\t (Q^\psi (\t)Q^\phi(\t))^2\\
    &= -\frac{((p-2)_2\,g^{p-3})^2}{2(p-1)g^{p-2}}\b\int d\t (Q^\psi(\t) Q^\phi(\t))^2
\end{split}
\ee 
Now the bosonic cubic terms are obtained from
\be 
\begin{split}
    &(p-2)_2\,g^{p-3}\b \int d\t \Big[ -\half(Q^\phi)^2 g_F + ((Q^\psi)^2 - Q^F Q^\phi)g_\phi \Big]\\
    =& -\frac{(p-2)_2\,g^{p-3}}{2} \sum_{\w \neq 0} \Big[ (2Q^F Q^\phi - 2(Q^\psi)^2 + \w^2 (Q^\phi)^2)(\w)\frac{\rho(\w)}{|\w|}\\
    &\hspace{120pt}+ (2Q^F Q^\phi - 2(Q^\psi)^2 - \w^2 (Q^\phi)^2)(\w)\frac{\gamma(\w)}{|\w|} \Big]
\end{split}
\ee 
The products among $Q_A$ above are taken in position space, then Fourier transformed to frequency space. Now the term accompanying $\gamma$ is proportional to 
\be 
(2Q^F Q^\phi - 2(Q^\psi)^2 + \del_\t^2 (Q^\phi)^2)(\w)\frac{\gamma(\w)}{|\w|} = (-2\del_\t^2 Q^\phi Q^\phi - 2(\del_\t Q^\phi)^2 + \del_\t^2 (Q^\phi)^2)(\w)\frac{\gamma(\w)}{|\w|} = 0
\ee 
Hence there is no linear interaction term for $\gamma$, and its equation of motion is $\gamma =0$. As a result, it doesn't contribute to the index. On the other hand, the $\rho$ interaction term can be written as 
\be 
\begin{split}
    &-\frac{(p-2)_2\,g^{p-3}}{2} \sum_{\w \neq 0} (2Q^F Q^\phi - 2(Q^\psi)^2 + \w^2 (Q^\phi)^2)(\w)\frac{\rho(\w)}{|\w|}\\
    =&-\frac{(p-2)_2\,g^{p-3}}{2} \sum_{\w \neq 0} (-2\del_\t^2 Q^\phi Q^\phi - 2(\del_\t Q^\phi)^2 -\del_\t^2 (Q^\phi)^2)(\w)\frac{\rho(\w)}{|\w|}\\
    =&-\frac{(p-2)_2\,g^{p-3}}{2} \sum_{\w \neq 0} -4\del_t(\del_t Q^\phi Q^\phi)(\w)\frac{\rho(\w)}{|\w|}\\
    =& 2i(p-2)_2\,g^{p-3} \sum_{\w \neq 0}(\del_\t Q^\phi Q^\phi)(\w)\frac{\w}{|\w|}\rho(\w)
\end{split}
\ee 
Now we can integrate out $\rho$ classically,  and the contribution to the index is 
\be 
\frac{((p-2)_2\,g^{p-3})^2}{2(p-1)g^{p-2}}\b \int d\t (Q^\phi (\t)\del_\t Q^\phi(\t))^2 = \frac{((p-2)_2\,g^{p-3})^2}{2(p-1)g^{p-2}}\b \int d\t (Q^\phi(\t) Q^\psi(\t))^2
\ee 
Note that it exactly cancels against (\ref{Iferm}), so the fluctuation of the cubic terms don't give any contribution. We conclude that our effective action (\ref{effA}) does not get any corrections up to the order we are interested in.

\textbf{Corrections to zero frequency modes:}   For the fermionic mode we write $G_\psi(\w=0)=\b G^{av}_\psi+g_\psi(\w=0)$. Just like the non-zero frequency counterpart, the quadratic piece in $g_\psi(\w=0)$ scales as $g^{2-p}$. Hence the only linear terms that can contribute are from sub-leading quadratic piece up to the order we are considering by the same argument as presented before. However the linear piece from sub-leading quadratic piece vanishes and hence on-shell  $g_\psi(\w=0)=0$. Finally the correction to $G_F(\w=0)=\b G^{av}_F+g_F(\w=0)$ is sub-leading because its quadratic piece scales as $g^{2(p-1)}>g^{p-2}$.  

This completes our argument for recovering the Witten index from bi-local path integral.

\subsection{Two copies of $\mathcal{N}=4$ supersymmetric SYK model}

In this sub-section we take two copies of $\mathcal{N}=4$ supersymmetric SYK model and call them left (L) and right (R) sub-systems. We can couple these two copies and still preserve diagonal super-symmetries that acts on both sides in the same way. One such coupling is given by 
\be
\int d\t(d^2\th \l_{ij}\Phi^i_L\Phi^j_R + d^2\bar{\th}\bar{\l}_{ij}\bPhi^i_L\bPhi^j_R )
\ee
The coupling also does not change the asymptotic nature of the super-potential for $p>2$. Hence even with this coupling the index of the two sided system will take the factorized form. We have verified this numerically. 
Our Mathematica code is attached as supplementary files to this paper.
The factorized index does not depend on the precise value of the SYK coupling or two sided coupling. This allows us to average over both of them to get a disordered theory. The novelty lies in recovering the exact value of the index from this disorder averaged coupled theory.  In fact we will see that the saddle points that contribute to the index comes from equations that couple $G^A_{LL}, G^A_{RR}$ ($A=\phi,\psi, F$). 

We average over the SYK coupling and $\lambda_{ij}$ with Gaussian measure:
\be 
 \langle \lambda_{i_1 i_2}\bar{\lambda}_{i'_1 i'_2} \rangle_D = \frac{\sigma^2}{ 2N}\delta_{i_1i'_1} \delta_{i_2 i'_2}
\ee 
With this interaction, the starting point for the large $N$ analysis should be  
\be 
\begin{split}
     \frac{S_{eff}}{N} &= - \log det\Big( -\de(\t - \t')\de_{IJ}\del^2_\t - \bar{\Si}^\phi_{IJ} \Big) -\log det\Big( \de(\t - \t')\de_{IJ} - \bar{\Si}^F_{IJ} \Big)\\
    &+ 2\log det\Big( \de(\t - \t')\de_{IJ}\del_\t - \bar{\Si}^\psi_{IJ} \Big)  - {\sigma^2}\int d\t d\t'\Big[ G^\phi_{LL}G^F_{RR} + G^F_{LL}G^\phi_{RR} - 2 G^\psi_{LL}G^\psi_{RR}\Big] \\
    & - \int d\t d\t' \Big(  \bar{\Si}^\phi_{IJ}G^\phi_{IJ} + \bar{\Si}^F_{IJ}G^F_{IJ} + 2\bar{\Si}^\psi_{IJ} G^\psi_{IJ} - (p-1)J^p (G^\psi_{IJ})^2(G^\phi_{IJ})^{p-2} + J^p G^F_{IJ}(G^\phi_{IJ})^2 \Big) 
\end{split}
\ee 
In frequency modes, it can be written as  
\be 
\begin{split}
    \frac{S_{eff}}{N} &= \sum_{\w} \Bigg[- \log\Big[ (\w^2 - \bar{\Si}^\phi_{LL})((\w^2 - \bar{\Si}^\phi_{RR})) - \bar{\Si}^\phi_{LR}\bar{\Si}^\phi_{RL} \Big] \\
    & - \log\Big[ (1 - \bar{\Si}^F_{LL})((1 - \bar{\Si}^F_{RR})) - \bar{\Si}^F_{LR}\bar{\Si}^F_{RL} \Big] \\
    & + 2\log\Big[ (-i\w - \bar{\Si}^\psi_{LL})((-i\w - \bar{\Si}^\psi_{RR})) - \bar{\Si}^\psi_{LR}\bar{\Si}^\psi_{RL} \Big] \\
    & -\bar{\Si}^\phi_{IJ}(\w)G^\phi_{IJ}(-\w) -\bar{\Si}^F_{IJ}(\w)G^F_{IJ}(-\w) + 2\bar{\Si}^\psi_{IJ}(\w)G^\psi_{IJ}(-\w)   \Bigg]\\
    & - \sum_{\w} \sigma^2 \Big[ G^\phi_{LL}(\w)G^F_{RR}(-\w) + G^\phi_{RR}(\w)G^F_{LL}(-\w) - 2G^\psi_{LL}(\w)G^\psi_{RR}(-\w) \Big]\\
    & + \int d\t d\t' \Big( (p-1)(G^\psi_{IJ}(\t-\t'))^2(G^\phi_{IJ}(\t-\t'))^{p-2} - G^F_{IJ}(\t-\t')(G^\phi_{IJ}(\t-\t'))^{p-1} \Big) 
\end{split}
\ee 
 Typical wormhole-like-saddles
have non-vanishing left right correlation, i.e., $G^A_{LR} \neq 0, \Si^A_{LR} \neq 0$ \cite{saad2019semiclassicalrampsykgravity}. In $\mathcal{N}=4$ SYK model that we consider here, we don't expect such contribution to be present. Therefore we look for saddles that satisfy $G^A_{LR} = \Si^A_{LR} =0$. From the equation above we note that, saddle point equation obtained by varying $G^A_{LR}, \Si^A_{LR}$ are automatically satisfied if we set $G^A_{LR} = \Si^A_{LR} =0$. This explains why this is a consistent background. On the other hand, in  $\mathcal{N}=2$ SYK model these wormhole like saddles might contribute to the disorder averaged index. We explore this very interesting possibility in our upcoming future work \cite{Halder2025}.  In this paper, we are looking for solutions that preserve diagonal supersymmetry. Hence, it is natural to look for a symmetric saddle point\footnote{Note that positivity constraints force us to have $G^{av,\phi}_{RR} \geq 0, G^{av,\phi}_{LL} \geq 0$. So it makes sense to work with the symmetric saddle point $G^\phi_{RR}=G^\phi_{LL}$. Relative sign of $G^\psi_{RR}, G^F_{RR}$ to $G^\phi_{RR}$ is determined by supersymmetry (same statement holds for left copy). Hence we look for the symmetric solution for all the components. } 
\be
G^A_{LL}= G^A_{RR}
\ee
As before we  assume non-zero frequency parts are small compared to the zero frequency parts of $G^A$. Define $G^{av, \phi}_{LL} = G^{av, \phi}_{RR} =g$. After expanding, we work up to leading quadratic order in non-zero frequency parts. This gives
\be 
\begin{split}
    \frac{S_{eff}}{N} &= \sum_{\w} \Bigg[- \log\Big[ (\w^2 - \bar{\Si}^\phi_{LL})((\w^2 - \bar{\Si}^\phi_{RR})) \Big] - \log\Big[ (1 - \bar{\Si}^F_{LL})((1 - \bar{\Si}^F_{RR})) \Big] \\
    & + 2\log\Big[ (-i\w - \bar{\Si}^\psi_{LL})((-i\w - \bar{\Si}^\psi_{RR})) \Big] \\
    & -\bar{\Si}^\phi_{LL}(\w)G^\phi_{LL}(-\w) -\bar{\Si}^F_{LL}(\w)G^F_{LL}(-\w) + 2\bar{\Si}^\psi_{LL}(\w)G^\psi_{LL}(-\w)\\
    & -\bar{\Si}^\phi_{RR}(\w)G^\phi_{RR}(-\w) -\bar{\Si}^F_{RR}(\w)G^F_{RR}(-\w) + 2\bar{\Si}^\psi_{RR}(\w)G^\psi_{RR}(-\w) \Bigg]\\
    & - \sum_{\w} \sigma^2 \Big[ G^\phi_{LL}(\w)G^F_{RR}(-\w) + G^\phi_{RR}(\w)G^F_{LL}(-\w) - 2G^\psi_{LL}(\w)G^\psi_{RR}(-\w) \Big]\\
    & - \b G_{LL}^F(\w=0)g^{p-1} + (p-1)g^{p-2} G^\psi_{LL}(\w=0)^2 \\
    &+ (p-1)g^{p-2}\sum_{\w \neq 0} \Big((\de G^\psi_{LL}(\w)\de G^\psi_{LL}(-\w)) - \de G^F_{LL}(\w)\de G^\phi_{LL}(-\w)\Big)\\
    & - \b G_{RR}^F(\w=0)g^{p-1} + (p-1)g^{p-2} G^\psi_{RR}(\w=0)^2\\
    &+ (p-1)g^{p-2}\sum_{\w \neq 0} \Big((\de G^\psi_{RR}(\w) \de G^\psi_{RR}(-\w)) - \de G^F_{RR}(\w) \de G^\phi_{RR}(-\w)\Big)
\end{split}
\ee 
  This is exactly 2 copies of the one sided theory, with the additional  $\sigma^2$ term. So we can follow the one-sided calculation to substitute the $\bar{\Si}$ saddles and get the effective action in terms of $G^A$ fields, which is 
\be 
\begin{split}\label{cIndex}
    \frac{S_{eff}}{N} &= -\b G^F_{LL}(\w=0)g^{p-1} + (p-1)g^{p-2}(G^\psi_{LL}(\w=0))^2 \\
    &\hspace{0pt} + \sum_{\w} \left[ \log(G^\phi_{LL}(\w)G^F_{LL}(\w)) - 2\log(G^\psi_{LL}(\w)) \right]+\sum_{\w}\Big[ -2i\w G^\psi_{LL}(\w) - \w^2G^\phi_{LL}(\w) - G^F_{LL}(\w)  \Big]\\
    &+ (p-1)g^{p-2}\sum_{\w \neq 0} \Big[ -\de G^F_{LL}(-\w) \de G^\phi_{LL}(\w) + \de G^\psi_{LL}(-\w) \de G^\psi_{LL}(\w)  \Big]\\
    & -\b G^F_{RR}(\w=0)g^{p-1} + (p-1)g^{p-2} (G^\psi_{RR})^2(\w=0)\\
    &\hspace{0pt}  + \sum_{\w} \left[\log(G^\phi_{RR}(\w)G^F_{RR}(\w)) - 2\log(G^\psi_{RR}(\w))\right]+\sum_{\w}\Big[ -2i\w G^\psi_{RR}(\w) - \w^2G^\phi_{RR}(\w) - G^F_{RR}(\w)  \Big] \\
    &+ (p-1)g^{p-2}\sum_{\w \neq 0} \Big[ -\de G^F_{RR}(-\w)\de G^\phi_{RR}(\w) + \de G^\psi_{RR}(-\w) \de G^\psi_{RR}(\w)  \Big]\\
    & +\sigma^2 \sum_{\w}  \Big[ -G^F_{RR}(-\w)G^\phi_{LL}(\w) - G^F_{LL}(-\w)G^\phi_{RR}(\w) +G^\psi_{RR}(-\w)G^\psi_{LL}(\w)+ G^\psi_{LL}(-\w)G^\psi_{RR}(\w)\Big]
\end{split}
\ee 
Equations of motion couple left and right sub-systems  due to the $\sigma^2$ term. This is an interesting situation where left right solutions are coupled even though left-right direct correlation is absent. 

\textbf{Zero frequency modes:} Equations of motion for zero frequency modes are given by 
\be 
\begin{split}
    G^F_{LL}(\w=0) &= \inv{1 + \b(G^{av,\phi}_{LL})^{p-1} + \sigma^2\b G^{av,\phi}_{RR}}\ , \ G^F_{RR}(\w=0) = \inv{1 + \b(G^{av,\phi}_{RR})^{p-1} + \sigma^2\b G^{av,\phi}_{LL}}\\
    &= \inv{1 + \b g^{p-1} + \sigma^2\b g} \hspace{110pt}=\inv{1 + \b g^{p-1} + \sigma^2\b g}
\end{split}
\ee 
and 
\be 
& (p-1)g^{p-2} G^\psi_{LL}(\w=0) - \inv{G^\psi_{LL}(\w=0)} + \sigma^2 G^\psi_{RR}(\w=0)=0 \\
& (p-1)g^{p-2} G^\psi_{RR}(\w=0) - \inv{G^\psi_{RR}(\w=0)} + \sigma^2 G^\psi_{LL}(\w=0)=0
\ee 
This is a quadratic equation, we pick the symmetric solution for the fermionic zero frequency modes  
\be 
G^\psi_{LL} (\w=0)=G^\psi_{RR}(\w=0)=  \inv{\sqrt{(p-1)g^{p-2} + \sigma^2}} \
\ee
Before discussing the nonzero frequency solutions, we check the contribution to the index from the zero modes (later we will see that the nonzero mode contributions cancel out, like before)
\be \label{zeromodeindex}
\begin{split}
    \frac{S_{eff}}{N} &\supset -1 - 1 + 2 + \log(\frac{\b g}{1 + \b g^{p-1} + \sigma^2 \b g}) + \log((p-1)g^{p-2} + \sigma^2) \\
    &\hspace{30pt}+ \log(\frac{\b g}{1 + \b g^{p-1} + \sigma^2 \b g}) + \log((p-1)g^{p-2} + \sigma^2)
\end{split}
\ee 
So the zero mode contribution to the index is given by  
\be 
\frac{\log\mathcal{I}}{N} \supset 2\log(p-1)+ 2\log\Big[ \frac{\b g^{p-1} + \sigma^2 \b g}{1 + \b g^{p-1} + \sigma^2\b g} \Big]
\ee 
Again we see that the contribution of zero frequency part to the index gets maximized at $g\to \infty$. Setting this saddle point value gives us the exact index. Essentially we see that the effect of $\sigma^2$ is not important when $g\to \infty$, which resonates well with the fact that this left-right coupling does not change the asymptotic nature of the super-potential. We expect this property to be general feature of such deformations. Now we turn to show that indeed the contribution of the non-zero modes to the index vanish.  

\textbf{Non-zero frequency modes:} The equations of motion for $\w>0$ modes of $G^A$ are 
\be 
\begin{split}
    &\inv{G^F_{LL}(\w)} - 1 - \sigma^2 G^\phi_{RR}(-\w) - (p-1)g^{p-2} G^\phi_{LL}(-\w)=0\\
    &\inv{G^F_{RR}(\w)} - 1 - \sigma^2 G^\phi_{LL}(-\w) - (p-1)g^{p-2} G^\phi_{RR}(-\w)=0\\
    &\inv{G^\phi_{LL}(\w)} - \w^2 - \sigma^2 G^F_{RR}(-\w) - (p-1)g^{p-2}G^F_{LL}(-\w)=0\\
    &\inv{G^\phi_{RR}(\w)} - \w^2 - \sigma^2 G^F_{LL}(-\w) - (p-1)g^{p-2} G^F_{RR}(-\w)=0\\
    &-\inv{G^\psi_{LL}(\w)} - i\w + \sigma^2 G^\psi_{RR}(-\w) + (p-1)g^{p-2} G^\psi_{LL}(-\w)=0\\
    &-\inv{G^\psi_{RR}(\w)} - i\w + \sigma^2 G^\psi_{LL}(-\w) + (p-1)g^{p-2} G^\psi_{RR}(-\w)=0
\end{split}
\ee 
The solution to these, compatible with the SUSY conditions, are given by
\be 
\begin{split}
    G^\psi_{LL}(\w) = G^\psi_{RR}(\w) = i sgn(\w) \frac{-|\w| + \sqrt{\w^2 + 4((p-1)g^{p-2} + \sigma^2)}}{2((p-1)g^{p-2} + \sigma^2)}\\
    G^F_{LL}(\w) = G^F_{RR}(\w) = |\w| \frac{-|\w| + \sqrt{\w^2 + 4((p-1)g^{p-2} + \sigma^2)}}{2((p-1)g^{p-2}+ \sigma^2)}\\
    G^\phi_{LL}(\w) = G^\phi_{RR}(\w) = \inv{|\w|} \frac{-|\w| + \sqrt{\w^2 + 4((p-1)g^{p-2} + \sigma^2)}}{2((p-1)g^{p-2}+ \sigma^2)}
\end{split}
\ee 
This is the same  as the solution for decoupled equations with the shift  $g^{p-2} \to g^{p-2} + \frac{\sigma^2}{p-1}$. It is easy to see from the explicit expression (\ref{cIndex}) that the nonzero frequency part of the coupled index  is just given by the shift $$g^{p-2} \to g^{p-2} + \frac{\sigma^2}{p-1}$$ to the decoupled index, and therefore the nonzero frequency contribution  cancels out, similar to earlier calculations.

\section{Discussion and future directions}

In this paper we have focused on the $\mathcal{N}=4$ supersymmetric SYK model and performed a disorder average on coupling constants that do not alter the value of the index. We also discussed saddle point solutions with non-trivial coupling between two copies of the theory for a fixed value of the zero mode of bi-local fields. These coupled solutions notably have vanishing left-right two point correlation function. The action for the bi-local fields still couples the left-left and right-right ones. The large $N$ saddle at $g\rightarrow \infty$ effectively decouples the system completely. At the present stage we lack understanding of a gravitational dual of these configurations where the zero mode of bi-local fields are off-shell and the action remains coupled.   Do they correspond to horizon-less configurations with huge entropy? It would be fantastic to make progress in this direction in the future.

Our arguments regarding the disorder-averaged representation for products of the Witten index are general and apply to any quantum field theory. Given the close similarity between the model discussed here and the BFSS model (see \cite{maldacena2023simplequantumdescribesblack, Lin_2023, Lin:2024vvg} and references therein for a recent review of the BFSS model), understanding such disorder average in BFSS model is an intriguing research direction. In addition, these techniques can be generalized to higher spacetime dimensions easily.

Here, we have limited our analysis to a specific class of couplings that preserve the value of the Witten index. Nonetheless, it is also possible to perform disorder averaging over supersymmetry-preserving coupling constants in a more general context. In such cases, the resulting quantity does not necessarily factorize across different copies of the theory, opening up the intriguing possibility of contributions from BPS analog of  Maldacena, Qi type wormhole configurations \cite{Chen:2023hra}. This suggests a richer structure in the disorder-averaged theories, which warrants further investigation. Some results along these directions will appear in an upcoming work by the authors \cite{Halder2025}.

\section{Acknowledgements}

We would like to thank Neeraj Tata for discussion in related projects. We thank H. Lin, J. Maldacena and D. Stanford for comments on a preliminary version of the manuscript. We also thank K. Budzik, Y. Chen, M. Heydeman, V. Hubeny, L. Iliesiu, V. Narovlansky, M. Rangamani, J. Trnka for helpful discussions. 
I. H. is supported by U.S. Department of Energy grant DE-SC0009999 and funds from the University of California.
The work of D.L.J. and D. S. is supported in part by  U.S. Department of Energy grant DE-SC0007870 and by the Simons Investigator in Physics Award MP-SIP-0001737. D.L.J. further thanks the Institute for Advanced Study for hospitality, and the IBM Einstein Fellow Fund for support. 

\appendix

\section{Standard method of evaluating the Witten index}\label{appW}
In this Appendix we discuss the standard method for evaluation of the Witten index for $\mathcal{N}=2$ quantum mechanics ($\mathcal{N}=4$ theories are to be considered as a special case). We start with the Euclidean action 
\be 
S =  \int d\t \Big[\half \psi^i_\al \dot{\psi}^i_\al + \half(F^iF^i) + \half(\dot{\phi}^i\dot{\phi}^i) + 3iC_{ijk}\big( F^i\phi^j\phi^k + \eps^{\al\b}\psi^i_\al\psi^j_\b\phi^k \big) \Big]
\ee 
and integrate out the auxiliary field $F^i$ to obtain 
\be 
S = \int d\t \Big[ \half \psi^i_\al \dot{\psi}^i_\al + \half(\dot{\phi}^i\dot{\phi}^i)  + 3iC_{ijk}\eps^{\al\b}\psi^i_\al\psi^j_\b\phi^k + \frac{1}{2}\sum_{i=1}^N \big[\del_i W(\phi)^2\big] \Big]
\ee 
here we defined the superpotential $W(\phi) = C_{ijk}\phi^i\phi^j\phi^k$. 
The Hamiltonian is given by
\be 
\begin{split}
H &= \half \pi^i\pi^i + \half \sum_i [\del_i W(\phi)^2] + 3iC_{ijk}\eps^{\al\b}\psi^i_\al\psi^j_\b\phi^k\\
&\equiv \half \pi^i\pi^i + \half \sum_i [\del_i W(\phi)^2] + \frac{i}{2}\del_i\del_jW(\phi)\eps^{\al\b}\psi^i_\al\psi^j_\b
\end{split}
\ee 
where $\pi^i = \frac{\del \mathcal{L}}{\del \dot{\phi}^i}$ is the canonical momentum for the boson. We quantize the theory and work in the usual position basis for the bosons : 
\be
\langle \boldsymbol{\tilde{\phi}} |\boldsymbol{\phi}\rangle=\delta^{(N)}(\tilde{\phi}-\phi), \quad \langle \boldsymbol{\phi} |\boldsymbol{\pi}\rangle= \frac{1}{(2\pi)^{\frac{N}{2}}}e^{i \boldsymbol{\phi}\cdot \boldsymbol{\pi}}
\ee
The fermionic Hilbert space is described as follows.

The canonical momenta of the Majorana fermion system is 
\be
\Pi^i_\al = \half \psi^i_\al
\ee 
and we have "second class primary constraints"
\be 
\chi^i_\al = \Pi^i_\al - \half \psi^i_\al=0
\ee 
with Poisson anti-bracket
\be 
K^{ij}_{\al,\b} = \{\chi^i_\al, \chi^j_\b\}^P = \de_{ij}\de_{\al\b}
\ee 
So the Dirac antibracket is defined as 
\be 
\{\xi, \eta \}^D = \{\xi, \eta\}^P- \{\xi, \chi^i_\al\}^P(K^{-1})^{\al\b}_{ij}\{\chi^j_\b, \eta\}^P
\ee 
Promoting the Dirac antibracket to anti-commutation relation between operators, we get
\be 
\{\psi^i_\al, \psi^j_\b\} = \de^{ij}\de_{\al\b}.
\ee 
Specifically
\be 
(\psi^i_1)^2 = (\psi^i_2)^2 = \half
\ee 
Starting with 2 real fermions for each $i$, we can create a pair of complex fermions 
\be 
\psi^i = \inv{\sqrt{2}}(\psi^i_1 - i\psi^i_2) \ ; \ \bpsi^i = \inv{\sqrt{2}}(\psi^i_1 + i\psi^i_2)
\ee 
which satisfy standard creation-annihilation operator algebra
\be 
\{\psi^i, \psi^j\} = \{\bpsi^i, \bpsi^j\} = 0 \ ; \ \{ \psi^i, \bpsi^j\}= \de^{ij}
\ee 
So the fermion part of the Hilbert space will consist of two level systems, for each $i$. We have an oscillator ground state $\ket{0}$ satisfying 
\be 
\psi^i\ket{0}=0, \ \ \forall i = 1, 2, \dots N
\ee 
The rest of states are of the form
\be \label{fermHil}
(\bpsi^N)^{n_N}\dots (\bpsi^1)^{n_1}\ket{0} \equiv \ket{n_N, \dots, n_1}
\ee 
Where $n_i \in \{0,1\}$. We can further write the potential term involving fermions in terms of creation-annihilation operators
\be 
-\frac{i}{2}\del_i\del_jW(\phi)\eps^{\al\b}\psi^i_\al\psi^j_\b = \frac{i}{2}\del_i\del_jW(\phi)(\psi^i_1\psi^j_2 - \psi^i_2\psi^j_1)= -\half\del_i\del_jW(\phi)[\bpsi^i, \psi^j] 
\ee 
Now we turn to evaluate the Witten index of the quantum mechanics:
\be 
\label{bosint2}
Tr(-1)^Fe^{-\b H}
\ee 
We will first evaluate the trace over the fermionic part of the Hilbert-space and then over the bosonic part in the limit $\beta \to 0+$ and finally use the fact that it is independent of $\beta$ for $H$ with a discrete spectrum to get an explicit formula in terms of the superpotential. We compute the trace over the fermion Hilbert space first, 
\be 
\begin{split}
    &Tr_{\phi}Tr^{(N)}_\psi (-1)^Fe^{-\b H}\\
    =&Tr_{\phi}\Bigg( \exp\Big(-\frac{\b}{2}(\pi^i\pi^i + [\del_iW(\phi)]^2)\Big) Tr^{(N)}_\psi (-1)^F\exp\Big( -\frac{\b}{2}\del_i\del_jW(\phi)[\bpsi^i, \psi^j] \Big)\Bigg)\\
    =&Tr_{\phi}\Bigg(\exp\Big(-\frac{\b}{2}(\pi^i\pi^i + [\del_iW(\phi)]^2)\Big) \sum_{\{n_i\}} \bra{\{n_i\}}(-1)^F \exp\big(-\frac{\b}{2}\del_i\del_jW(\phi)[\bpsi^i, \psi^j]\big)\ket{\{n_i\}}\Bigg)
\end{split}
\ee 
, where $(N)$ denotes trace over $N$ complex fermion Hilbert space. For a fixed value of $\phi$, $\del_i\del_jW(\phi)$ is just a $N\times N$ matrix, which can be diagonalised. By doing the appropriate similarity transform, and rotating the fermions accordingly, we can write the expression as 
\be \label{ferm2}
Tr^{(N)}_{\tilde{\psi}} (-1)^F \exp\Big[-\sum_k \frac{\b}{2} \l_k (\tilde{\bpsi}^k \tilde{\psi}^k - \tilde{\psi}^k\tilde{\bpsi}^k) \Big]
\ee 
Where $\del_i\del_j W(\phi) \to diag(\{\l_k(\phi)\}), (\bpsi^i, \psi^i) \to (\tilde{\bpsi}^k, \tilde{\psi}^k)$ under the diagonalisation. Now we can take the trace over the Hilbert space where the $\tilde{\psi}^k$ and $\tilde{\bpsi}^k$ act as annihilation and creation operators. Furthermore the operator $\hat{N}_k = [\tilde{\bpsi}^k, \tilde{\psi}^k]$ is the fermion number operator, $\hat{N}_k : \ket{0}\to -1, \tilde{\bpsi}_k\ket{0}\to +1$. Because of the diagonal structure, we can look at individual fermions independently and write (\ref{ferm2}) as
\be 
\prod_k Tr^{(1)}_{\psi} (-1)^F \exp\Big( -\frac{\b}{2} \l_k [\tilde{\bpsi}^k, \tilde{\psi}^k] \Big) = \prod_k Tr^{(1)} (-1)^F \exp(-\frac{\b}{2}\l_k \hat{N}_k)
\ee 
The individual traces are over single complex fermion Hilbert spaces, which is easy to compute, 
\be 
\begin{split}
    &\prod_k Tr^{(1)}_{\psi} (-1)^F \exp\Big[ -\b \l_k \hat{N}_k \Big] = \prod_k  \Big( \exp[\b\l_k] + (-1)\exp[-\b\l_k]\Big)\\
    =& \prod_k (2sinh(\frac{\b}{2}\l_k)) \xrightarrow{\b\to 0^{+}} \prod_k (\b\l_k)= \b^N det(\del_i \del_j W(\phi))
\end{split}
\ee 
Thus, at high temperature, the index takes the form
\be \label{onlyboson}
\lim_{\b\to 0+}Tr(-1)^Fe^{-\b H} = {\b^N}\int d^N\phi \bra{\boldsymbol{\phi}} \,det(\del_i\del_jW(\phi)) \exp\Big[ -\frac{\b}{2}(\pi^2 +\sum_i\del_iW(\phi)^2) \Big] \ket{\boldsymbol{\phi}}
\ee 
Since $\phi, \pi$ do not commute with each other we need to use the Baker–Campbell–Hausdorff formula to compute this quantity. However since we are only interested in the leading $\beta\to 0+$ dependence we can still factorize the exponential and use 
\be 
\bra{\boldsymbol{\tilde{\phi}}} \exp(-\frac{\b}{2}\pi^2)\ket{\boldsymbol{\phi}}=\int d^N \pi  \exp(-\frac{\b}{2}\pi^2) \langle \boldsymbol{\tilde{\phi}}| \boldsymbol{\pi} \rangle \langle \boldsymbol{\pi}| \boldsymbol{\phi}\rangle = \inv{(2\pi\b)^{N/2}}\exp\Big( - \frac{(\tilde{\phi} - \phi)^2}{2\b}\Big)
\ee 
Plugging this back in, we get
\be \label{bosint1}
\lim_{\b \to 0+}Tr(-1)^Fe^{-\b H} =\frac{\b^N}{(2\pi\b)^{N/2}}\int d^N\boldsymbol{\phi}  \,det(\del_i\del_jW(\phi)) \exp\Big[ -\frac{\b}{2} \sum_i\del_iW(\phi)^2 \Big] 
\ee 
It is convenient to define $\tilde{\phi}= \b^{-\half}\phi$. Then the index takes the form 
\be \label{bosint3}
\lim_{\b \to 0+}Tr(-1)^Fe^{-\b H} =\frac{1}{(2\pi)^{N/2}}\int d^N\boldsymbol{\tilde{\phi}}  \,det(\del_i\del_jw(\tilde{\phi})) \exp\Big[ -\frac{1}{2} \sum_i\del_iw(\tilde{\phi})^2 \Big] 
\ee 
Where we have made the following replacement
\be 
w(\tilde{\phi}) = W(\phi)\vert_{\phi \to \sqrt{\b} \tilde{\phi}} 
\ee 
In doing this replacement, the coefficients of the potential $w(\tilde{\phi})$ gets nontrivial $\b$ dependence.  Next, we change the variable of integration to $y_i = \tilde{\del}_i w(\tilde{\phi})$, which absorbs the determinant as the Jacobian.  We can write down the index as 
\be 
\begin{split}
\lim_{\b \to 0+}Tr(-1)^Fe^{-\b H} &= \Omega_Y = \inv{(2\pi)^{N/2}}\int_Y \prod_i \Big(dy_i e^{-\frac{y_i^2}{2}} \Big)
\end{split}
\ee 
where the integration domain is given by 
\be 
Y = \cup_\Delta Y_\Delta \  ,\, Y_\Delta = \{ \tilde{\del}_iw(\tilde{\phi})\,\vert \ \tilde{\phi}^i \in \Delta \subset \mathbb{R}^N \} 
\ee 
where we have defined $\Delta$ such that $\mathbb{R}^N = \cup\Delta$, and each $\Delta$ is the maximal set in $\mathbb{R}^N$ that induces a unique orientation in the target $Y_\Delta$ under the map $\tilde{\phi}^i \to y_i = \tilde{\del}_i(w(\tilde{\phi}))$. \\
We focus on theories with discrete spectrum and scale the potential with a positive constant $w(\tilde{\phi})\to C\cdot w(\tilde{\phi})$ to obtain
\be 
\mathcal{I} = \lim_{C\to +\infty} \Omega_{CY}
\ee 
From the above expression, we can see that for large $C$, the Gaussian integrals will localise near $y_i = 0$, i.e. $\del_i w(\tilde{\phi_*})=0$. Gaussian integral in each $\Delta$ will give a contribution $(-1)^{\gamma(\tilde{\phi}_*)}$, where $\gamma(\tilde{\phi}_*)$ is the Morse index of $\tilde{\phi}_*$, i.e. the number of negative eigenvalues of $\tilde{\del}_i\tilde{\del}_j w(\tilde{\phi_*})$, or equivalently of $\del_i\del_jW(\phi_*) $. So we obtain the index 
\be 
\begin{split}
    \mathcal{I}=&  \sum_{\{\phi_*\}} sgn\Big(det(\del_i\del_jW(\phi_*))\Big)
\end{split}
\ee 

Here we have assumed $\del_i\del_j W(\phi)$ is non-degenerate at the saddle points $\{\phi_*\}$. For the example of $\mathcal{N}=2$ SYK model we achieve this by deforming the super potential to
\be
W(\Phi) \to W(\Phi)+v_i\Phi^i
\ee

\section{Divergence from disorder averaging and its resolution}\label{appA}

In the main text we have studied the leading order large $N$ limit of the index in detail. The goal of this appendix is to perform an exact in $N$ analysis of the zero mode sector of the index $\mathcal{I}_0$. This will teach us qualitative facts about what to expect for the complete index. After disorder averaging the index takes the following form 
\be 
\mathcal{I}_0 \propto \int d[\phi, \psi, F] e^{-\bF F(1 + \text{number }\times |\phi|^{2p-2}) + \text{fermionic terms} }
\ee 
Integrating out the fermions and $F, \bar{F}$ gives power law in $\phi$ to the integrand, i.e., after other zero modes are integrated out , the integrand for $G_\phi^{av}$ lacks an exponential suppression for the  for large values as expected from a classical potential and  
hence, controlled by sensitive loop effects.
In particular, we show that unless certain `off diagonal' terms are considered, these loop effects are such that the integrand for $G_\phi^{av}$ has a local maxima at a finite value, which in leading order in $N$ diverges, and the integral representing the zero mode contribution to the index has a logarithmic divergence from large values of  $G_\phi^{av}$. More specifically,
\be 
\begin{split}
    \mathcal{I}_{0}^{\text{diag}} \propto (p-1)^N\int \frac{d\phi^id\bphi^i}{N} \frac{|\phi/\sqrt{N}|^{(2p-4)N}}{(1 + \text{number }\times|\phi/\sqrt{N}|^{2p-2})^N} 
\end{split}
\ee 
To understand this divergence better we restrict our consideration to $N=1$ and show that the `off diagonal' terms in effect regulates the divergence in the index giving it a finite value. However, the disorder averaged value of higher moments of $G_\phi^{av}$ diverge even when these off-diagonal effects are included.

\subsection*{Zero mode index for $N=1$}
 The zero mode index for $N=1$ is given by  
\be \label{N1index}
\begin{split}
    \mathcal{I}_0 &= \inv{(2\pi)^2}\int d[\phi, F, \psi_+, \psi_-]\exp\Big[-\Big(F\bar{F} + piC(F\phi^{p-1} + \frac{p-1}{2}(\psi_+\psi_- - \psi_-\psi_+)\phi^{p-2})\\
    &\hspace{100pt}+ pi\bar{C}(\bF\bphi^{p-1} + \frac{p-1}{2}(\bpsi_+\bpsi_- - \bpsi_-\bpsi_+)\bphi^{p-2})\Big)\Big]
\end{split}
\ee 
where $C$ is a complex random coupling and we have chosen Gaussian normalization for the bosonic fields, and unity for the fermions.

\subsubsection*{Computation without disorder averaging}
We can keep the coupling $C$ fixed and compute the index. First we can integrate out the fermions to get
\be 
\begin{split}
    \mathcal{I}_0 &= \inv{(2\pi)^2}\int d[\phi, F] \exp\Big[ -(F + ip\bar{C}\bphi^{p-1})(\bF + ipC\phi^{p-1}) - p^2|C|^2|\phi|^{2p-2} \Big]\times p^2(p-1)^2|C|^2|\phi|^{2p-4}
\end{split}
\ee 
Here we also completed the square for the $F$ field, which we can integrate out to get 
\be \label{onlyboson1}
\mathcal{I}_0 = \inv{2\pi}\int d\phi d\bphi \exp\Big[ - p^2|C|^2|\phi|^{2p-2}\Big] (p(p-1))^2 |C|^2 |\phi|^{2p-4}
\ee 
This integral can be explicitly carried out, and give us the answer 
\be 
\mathcal{I}_0 = p-1
\ee 
Interestingly this is the exact value of the index - it is related to computing the index in high temperature limit. It is independent of $C$.
 
 Moreover, we can also compute the expectation value of $G^{av}_\phi=|\phi|^2$ using path integral description. It is finite for generic coupling $C$ due to the exponential suppression in the effective measure. It scales with $C$ as $|C|^{-2/(p-2)}$, hence when we average over $C$ with a Gaussian measure with vanishing mean, one finds that $(G^{av}_\phi)^n$  diverges for $p>2$ due to the contribution from small $|C|$ if $n \geq p-1$.

\subsection*{Computation after disorder averaging}
Disorder averaging the zero mode index over the coupling constant $C$ gives us the expression
\be 
\mathcal{I}_0 =\int \frac{d[\phi, \psi, F]}{(2\pi)^2} \exp\Big[-\Big( \bF F + p^2\s^2(F\phi^{p-1}  + (p-1)\psi_+\psi_ -\phi^{p-2})(\bF\bphi^{p-1}  + (p-1)\bpsi_+\bpsi_-\bphi^{p-2})\Big) \Big]
\ee 
where $\overline{C^2} = \s^2$. After this point, in our large $N$ analysis, we only kept certain diagonal terms in the second product. The $N=1$ analogue of that calculation would look like the following
\be 
\mathcal{I}^{\text{diag}}_0 =\int \frac{d[\phi, \psi, F]}{(2\pi)^2} \exp\Big[-\Big( \bF F(1 + 9p^2\s^2|\phi|^{2p-2}) + p^2(p-1)^2\s^2|\phi|^{2p-4}\psi_+\psi_-\bpsi_+\bpsi_-\Big) \Big]
\ee 
where the diag superscript indicates that we have made the above mentioned assumptions. Then we can integrate out the fermions to get 
\be 
\begin{split}\label{divint}
  \mathcal{I}^{\text{diag}}_0 &= \inv{2\pi}\int d[\phi, F]\exp \Big[- \bF F (1 + p^2\s^2|\phi|^{2p-2})\Big]\times (p(p-1))^2\s^2|\phi|^{2p-4} \\
    &= \frac{p-1}{2\pi}\int d\phi d\bphi\ \frac{p^2(p-1)\s^2|\phi|^{2p-4}}{1 + p^2\s^2|\phi|^{2p-2}} 
\end{split}
\ee 
This integral is logarithmically divergent due to contribution from large $|\phi|$.  We can see that after disorder averaging, the effective suppression for the fields $\phi$ is rational, instead of exponential. This, together with the fermion contributions, makes the integral divergent. Note that the integrand has a maxima at a finite value. As mentioned previously, this local maxima moves away to infinity for large $N$. Furthermore, this also implies within this approximation that the expectation value of $|\phi|^2$ also diverges. Next we turn to show that the zero mode index becomes well defined when off-diagonal contributions are included.

\subsection*{Including off-diagonal terms}
Keeping off diagonal contributions, after disorder averaging the zero mode index becomes\footnote{Note that we have used the convention $\int d[\psi] \psi_+\psi_-\bpsi_+\bpsi_- = -1$.}
\be 
\begin{split}
\mathcal{I}_0 &=\int \frac{d[\phi, F, \psi]}{(2\pi)^2} \exp\Big[-\Big( \bF F + p^2\s^2(F\phi \phi + (p-1)\psi_+\psi_ -\phi)(\bF\bphi \bphi + (p-1)\bpsi_+\bpsi_-\bphi)\Big) \Big]\\
 &=\int \frac{d[\phi, F, \psi]}{(2\pi)^2} \exp\Big[-\Big( \bF F(1 + p^2\s^2|\phi|^{2p-2}) + (p(p-1))^2\s^2|\phi|^{2p-4}\psi_+\psi_-\bpsi_+\bpsi_- \\
 &\hspace{70pt}+ p^2(p-1)\s^2 F \phi |\phi|^{2p-4} \bpsi_+\bpsi_- + p^2(p-1)\s^2 \bF \bphi |\phi|^{2p-4} \psi_+\psi_- \Big)\Big]\\
 &= \int \frac{d[\phi, F, \psi]}{(2\pi)^2} \exp\Big[-\bF F(1 + p^2\s^2|\phi|^{2p-2})\Big](1- (p(p-1))^2\s^2|\phi|^{2p-4}\psi_+\psi_-\bpsi_+\bpsi_-)\\
 &\hspace{50pt}(1-p^2(p-1)\s^2 F \phi |\phi|^{2p-4} \bpsi_+\bpsi_-)(1-p^2(p-1)\s^2 \bF \bphi |\phi|^{2p-4} \psi_+\psi_-)\\
 &= \int \frac{d[\phi, F]}{(2\pi)^2} \exp\Big[-\bF F(1 + p^2\s^2|\phi|^{2p-2})\Big] \Big( (p(p-1))^2\s^2|\phi|^{2p-4} - (p^2(p-1))^2\s^4|F|^2|\phi|^{4p-6} \Big)
\end{split}
\ee 
So the effect of including the nondiagonal interactions is the second term above.  Now we perform the $F, \bF$ integral to get
\be \label{realbosint}
\begin{split}
\mathcal{I}_0 &= \inv{2\pi}\int d\phi d\bphi \frac{p^2(p-1)^2\s^2|\phi|^{2p-4}}{1+p^2\s^2|\phi|^{2p-2}} - \frac{ (p^2(p-1))^2 \s^4|\phi|^{4p-6}}{(1+p^2\s^2|\phi|^{2p-2})^2} = \inv{2\pi}\int d\phi d\bphi \frac{p^2(p-1)^2\s^2|\phi|^{2p-4} }{(1+p^2\s^2|\phi|^{2p-2})^2}
\end{split}
\ee 
This is a convergent integral, and can be computed as follows. We define
\be 
\phi = \inv{\sqrt{2}}(\phi_1 + i\phi_2) \ ; \ \bphi = \inv{\sqrt{2}}(\phi_1 - i\phi_2), 
\ee 
and switch to radial coordinates
\be \label{varchange}
(\phi_1, \phi_2) \to (r, \theta) \ ; \ |\phi|^2 \to \half r^2 \ ; \ d\phi d\bphi\to d\phi_1d\phi_2 \to rdrd\theta
\ee 
Now the integral can be done straightforwardly to give
\be 
\mathcal{I}_0 = (p-1)\int dr \frac{(2p-2)\l^2r^{2p-3}}{(1+ \l^2r^{2p-2})^2} = (p-1), \quad \l^2 = \frac{p^2\s^2}{2^{p-1}}
\ee 
This also implies the expectation value of
$(G^{av}_\phi)^n$  diverges for $p>2$ due to the contribution from large $r$ if $n \geq p-1$.

\providecommand{\href}[2]{#2}\begingroup\raggedright\endgroup


\begin{thebibliography}{10}

\bibitem{Strominger_1996}
A.~Strominger and C.~Vafa, {\it Microscopic origin of the bekenstein-hawking entropy},  {\em Physics Letters B} {\bf 379} (June, 1996) 99–104.

\bibitem{Iliesiu:2021are}
L.~V. Iliesiu, M.~Kologlu, and G.~J. Turiaci, {\it {Supersymmetric indices factorize}},  {\em JHEP} {\bf 05} (2023) 032, [\href{http://xxx.lanl.gov/abs/2107.09062}{{\tt arXiv:2107.09062}}].

\bibitem{Boruch:2023gfn}
J.~Boruch, L.~V. Iliesiu, S.~Murthy, and G.~J. Turiaci, {\it {New forms of attraction: Attractor saddles for the black hole index}},  [\href{http://xxx.lanl.gov/abs/2310.07763}{{\tt arXiv:2310.07763}}].

\bibitem{Maldacena_1999}
J.~Maldacena {\em International Journal of Theoretical Physics} {\bf 38} (1999), no.~4 1113–1133.

\bibitem{Lin:2022zxd}
H.~W. Lin, J.~Maldacena, L.~Rozenberg, and J.~Shan, {\it {Looking at supersymmetric black holes for a very long time}},  {\em SciPost Phys.} {\bf 14} (2023), no.~5 128, [\href{http://xxx.lanl.gov/abs/2207.00408}{{\tt arXiv:2207.00408}}].

\bibitem{Heydeman:2020hhw}
M.~Heydeman, L.~V. Iliesiu, G.~J. Turiaci, and W.~Zhao, {\it {The statistical mechanics of near-BPS black holes}},  {\em J. Phys. A} {\bf 55} (2022), no.~1 014004, [\href{http://xxx.lanl.gov/abs/2011.01953}{{\tt arXiv:2011.01953}}].

\bibitem{Turiaci:2023wrh}
G.~J. Turiaci, {\it {New insights on near-extremal black holes}},  [\href{http://xxx.lanl.gov/abs/2307.10423}{{\tt arXiv:2307.10423}}].

\bibitem{Halder:2023adw}
I.~Halder and D.~L. Jafferis, {\it {Thermal Bekenstein-Hawking entropy from the worldsheet}},  {\em JHEP} {\bf 05} (2024) 136, [\href{http://xxx.lanl.gov/abs/2310.02313}{{\tt arXiv:2310.02313}}].

\bibitem{Halder:2024gwe}
I.~Halder and D.~L. Jafferis, {\it {Stretched horizon, replica trick and off-shell winding condensate, and all that}},  [\href{http://xxx.lanl.gov/abs/2402.00932}{{\tt arXiv:2402.00932}}].

\bibitem{Ahmadain:2022tew}
A.~Ahmadain and A.~C. Wall, {\it {Off-shell strings I: S-matrix and action}},  {\em SciPost Phys.} {\bf 17} (2024), no.~1 005, [\href{http://xxx.lanl.gov/abs/2211.08607}{{\tt arXiv:2211.08607}}].

\bibitem{Ahmadain:2022eso}
A.~Ahmadain and A.~C. Wall, {\it {Off-shell strings II: Black hole entropy}},  {\em SciPost Phys.} {\bf 17} (2024), no.~1 006, [\href{http://xxx.lanl.gov/abs/2211.16448}{{\tt arXiv:2211.16448}}].

\bibitem{Ahmadain:2024hgd}
A.~Ahmadain and R.~Khan, {\it {A Worldsheet Derivation of the Classical Off-shell Boundary Action for the Dilaton in Half-Space}},  [\href{http://xxx.lanl.gov/abs/2406.00712}{{\tt arXiv:2406.00712}}].

\bibitem{Ahmadain:2024uyo}
A.~Ahmadain, S.~Akhtar, and R.~Khan, {\it {The GHY boundary term from the string worldsheet to linear order}},  [\href{http://xxx.lanl.gov/abs/2411.06400}{{\tt arXiv:2411.06400}}].

\bibitem{Firat:2024kxq}
A.~H. F\i{}rat and R.~A. Mamade, {\it {Boundary terms in string field theory}},  [\href{http://xxx.lanl.gov/abs/2411.16673}{{\tt arXiv:2411.16673}}].

\bibitem{penington2020entanglementwedgereconstructioninformation}
G.~Penington, {\it Entanglement wedge reconstruction and the information paradox},  2020.

\bibitem{Almheiri_2019}
A.~Almheiri, N.~Engelhardt, D.~Marolf, and H.~Maxfield, {\it The entropy of bulk quantum fields and the entanglement wedge of an evaporating black hole},  {\em Journal of High Energy Physics} {\bf 2019} (Dec., 2019).

\bibitem{brown2024evaporationchargedblackholes}
A.~R. Brown, L.~V. Iliesiu, G.~Penington, and M.~Usatyuk, {\it The evaporation of charged black holes},  2024.

\bibitem{biggs2025followingstateevaporatingcharged}
A.~Biggs, {\it Following the state of an evaporating charged black hole into the quantum gravity regime},  2025.

\bibitem{Almheiri_2020I}
A.~Almheiri, R.~Mahajan, J.~Maldacena, and Y.~Zhao, {\it The page curve of hawking radiation from semiclassical geometry},  {\em Journal of High Energy Physics} {\bf 2020} (Mar., 2020).

\bibitem{Almheiri_2020II}
A.~Almheiri, T.~Hartman, J.~Maldacena, E.~Shaghoulian, and A.~Tajdini, {\it Replica wormholes and the entropy of hawking radiation},  {\em Journal of High Energy Physics} {\bf 2020} (May, 2020).

\bibitem{penington2020replicawormholesblackhole}
G.~Penington, S.~H. Shenker, D.~Stanford, and Z.~Yang, {\it Replica wormholes and the black hole interior},  2020.

\bibitem{saad2019semiclassicalrampsykgravity}
P.~Saad, S.~H. Shenker, and D.~Stanford, {\it A semiclassical ramp in syk and in gravity},  2019.

\bibitem{Balasubramanian:2022gmo}
V.~Balasubramanian, A.~Lawrence, J.~M. Magan, and M.~Sasieta, {\it {Microscopic Origin of the Entropy of Black Holes in General Relativity}},  {\em Phys. Rev. X} {\bf 14} (2024), no.~1 011024, [\href{http://xxx.lanl.gov/abs/2212.02447}{{\tt arXiv:2212.02447}}].

\bibitem{akers2022blackholeinteriornonisometric}
C.~Akers, N.~Engelhardt, D.~Harlow, G.~Penington, and S.~Vardhan, {\it The black hole interior from non-isometric codes and complexity},  2022.

\bibitem{Gao_2022}
P.~Gao, D.~L. Jafferis, and D.~K. Kolchmeyer, {\it An effective matrix model for dynamical end of the world branes in jackiw-teitelboim gravity},  {\em Journal of High Energy Physics} {\bf 2022} (Jan., 2022).

\bibitem{iliesiu2024nonperturbativebulkhilbertspace}
L.~V. Iliesiu, A.~Levine, H.~W. Lin, H.~Maxfield, and M.~Mezei, {\it On the non-perturbative bulk hilbert space of jt gravity},  2024.

\bibitem{lee2023tracerelationsopenstring}
J.~H. Lee, {\it Trace relations and open string vacua},  2023.

\bibitem{lee2024bulkthimblesdualtrace}
J.~H. Lee and D.~Stanford, {\it Bulk thimbles dual to trace relations},  2024.

\bibitem{batra2025giantgravitonsdpbraneholography}
G.~Batra and H.~Lin, {\it Giant gravitons in d$p$-brane holography},  2025.

\bibitem{maldacena2018eternaltraversablewormhole}
J.~Maldacena and X.-L. Qi, {\it Eternal traversable wormhole},  2018.

\bibitem{Gao_2017}
P.~Gao, D.~L. Jafferis, and A.~C. Wall, {\it Traversable wormholes via a double trace deformation},  {\em Journal of High Energy Physics} {\bf 2017} (Dec., 2017).

\bibitem{Anninos_2016}
D.~Anninos, T.~Anous, and F.~Denef, {\it Disordered quivers and cold horizons},  {\em Journal of High Energy Physics} {\bf 2016} (Dec., 2016).

\bibitem{Murugan:2017eto}
J.~Murugan, D.~Stanford, and E.~Witten, {\it {More on Supersymmetric and 2d Analogs of the SYK Model}},  {\em JHEP} {\bf 08} (2017) 146, [\href{http://xxx.lanl.gov/abs/1706.05362}{{\tt arXiv:1706.05362}}].

\bibitem{Biggs_2024}
A.~Biggs, J.~Maldacena, and V.~Narovlansky, {\it A supersymmetric syk model with a curious low energy behavior},  {\em Journal of High Energy Physics} {\bf 2024} (Aug., 2024).

\bibitem{Polchinski_2016}
J.~Polchinski and V.~Rosenhaus, {\it The spectrum in the sachdev-ye-kitaev model},  {\em Journal of High Energy Physics} {\bf 2016} (Apr., 2016) 1–25.

\bibitem{Maldacena_2016}
J.~Maldacena and D.~Stanford, {\it Remarks on the sachdev-ye-kitaev model},  {\em Physical Review D} {\bf 94} (Nov., 2016).

\bibitem{maldacena2016conformalsymmetrybreakingdimensional}
J.~Maldacena, D.~Stanford, and Z.~Yang, {\it Conformal symmetry and its breaking in two dimensional nearly anti-de-sitter space},  2016.

\bibitem{Fu_2017}
W.~Fu, D.~Gaiotto, J.~Maldacena, and S.~Sachdev, {\it Publisher’s note: Supersymmetric sachdev-ye-kitaev models [phys. rev. d 95 , 026009 (2017)]},  {\em Physical Review D} {\bf 95} (Mar., 2017).

\bibitem{Witten:2016iux}
E.~Witten, {\it {An SYK-Like Model Without Disorder}},  {\em J. Phys. A} {\bf 52} (2019), no.~47 474002, [\href{http://xxx.lanl.gov/abs/1610.09758}{{\tt arXiv:1610.09758}}].

\bibitem{Klebanov_2017}
I.~R. Klebanov and G.~Tarnopolsky, {\it Uncolored random tensors, melon diagrams, and the sachdev-ye-kitaev models},  {\em Physical Review D} {\bf 95} (Feb., 2017).

\bibitem{Choudhury:2017tax}
S.~Choudhury, A.~Dey, I.~Halder, L.~Janagal, S.~Minwalla, and R.~Poojary, {\it {Notes on melonic $O(N)^{q-1}$ tensor models}},  {\em JHEP} {\bf 06} (2018) 094, [\href{http://xxx.lanl.gov/abs/1707.09352}{{\tt arXiv:1707.09352}}].

\bibitem{auffinger2011randommatricescomplexityspin}
A.~Auffinger, G.~B. Arous, and J.~Cerny, {\it Random matrices and complexity of spin glasses},  2011.

\bibitem{em/1175789759}
D.~H. Bailey, J.~M. Borwein, and D.~M. Bradley, {\it {Experimental Determination of Apéry-like Identities for $\zeta(2n+2)$}},  {\em Experimental Mathematics} {\bf 15} (2006), no.~3 281 -- 290.

\bibitem{Halder2025}
I.~Halder, D.~Jafferis, and D.~Sarkar, ``{Work in progress}.''

\bibitem{maldacena2023simplequantumdescribesblack}
J.~Maldacena, {\it A simple quantum system that describes a black hole},  2023.

\bibitem{Lin_2023}
H.~W. Lin, {\it Bootstrap bounds on d0-brane quantum mechanics},  {\em Journal of High Energy Physics} {\bf 2023} (June, 2023).

\bibitem{Lin:2024vvg}
H.~W. Lin and Z.~Zheng, {\it {Bootstrapping ground state correlators in matrix theory. Part I}},  {\em JHEP} {\bf 01} (2025) 190, [\href{http://xxx.lanl.gov/abs/2410.14647}{{\tt arXiv:2410.14647}}].

\bibitem{Chen:2023hra}
Y.~Chen, V.~Ivo, and J.~Maldacena, {\it {Comments on the double cone wormhole}},  {\em JHEP} {\bf 04} (2024) 124, [\href{http://xxx.lanl.gov/abs/2310.11617}{{\tt arXiv:2310.11617}}].

\end{thebibliography}
\end{document}